\documentclass[prx,amsmath,amssymb,twocolumn,longbibliography]{revtex4-1}
\usepackage{graphicx}
\usepackage{dcolumn}
\usepackage{bm}
\usepackage[utf8]{inputenc}
\usepackage[english]{babel}
\usepackage{amsfonts}
\usepackage{hyperref}
\usepackage{physics}
\usepackage{epstopdf}
\usepackage{mathtools}
\usepackage[capitalise]{cleveref}
\usepackage{float}
\usepackage[space]{grffile}
\usepackage{color}
\usepackage[normalem]{ulem}
\usepackage{subfigure}
\usepackage{natbib}
\usepackage[normalem]{ulem}
\usepackage[export]{adjustbox}

\binoppenalty=\maxdimen
\relpenalty=\maxdimen

\addto\captionsenglish{}

\def \hrho{\hat{\rho}}

\def \hc{\hat{c}}

\def \hH{\hat{H}}

\def \Heff{\boldsymbol{H}_{\rm eff}}

\begin{document}
	
	\title{Non-equilibrium stationary states of quantum non-Hermitian lattice models}
	
	\author{A. McDonald$^{1,2}$, R. Hanai$^{1,3,4}$ and A. A. Clerk$^1$}
	\affiliation{$^1$Pritzker School of Molecular Engineering, University of Chicago, Chicago, IL 60637, USA\\
	$^2$Department of Physics, University of Chicago, Chicago, IL 60637, USA\\
	$^3$Asia Pacific Center for Theoretical Physics, Pohang 37673, Korea \\
	$^4$Department of Physics, POSTECH, Pohang 37673, Korea}
	

	\begin{abstract}
	We show how generic non-Hermitian tight-binding lattice models can be realized in an unconditional, quantum-mechanically consistent manner by constructing an appropriate open quantum system.
	We focus on the quantum steady states of such models for both fermionic and bosonic systems.   Surprisingly, key features and spatial structures in the steady state cannot be simply understood from the non-Hermitian Hamiltonian alone. Using the 1D Hatano-Nelson model as a paradigmatic example, we show that the steady state has a marked sensitivity to boundary conditions.  In particular, the open boundary system can exhibit a large macroscopic length scale,
	despite having no corresponding long timescale.
	These effects persist in more general models, and are distinct from the localization physics associated with the non-Hermitian skin effect.
	Further,  particle statistics play an unexpected role:  the  
	steady-state density profile is dramatically different for fermions versus bosons.  Our work highlights the key role of fluctuations in quantum realizations of non-Hermitian dynamics, and provides a starting point for future work on engineered steady states of open quantum systems.
	
%
	\end{abstract}

	\maketitle
	

	\begin{figure*}[t]
	\centering
	\includegraphics[
	width=0.95\textwidth
	]{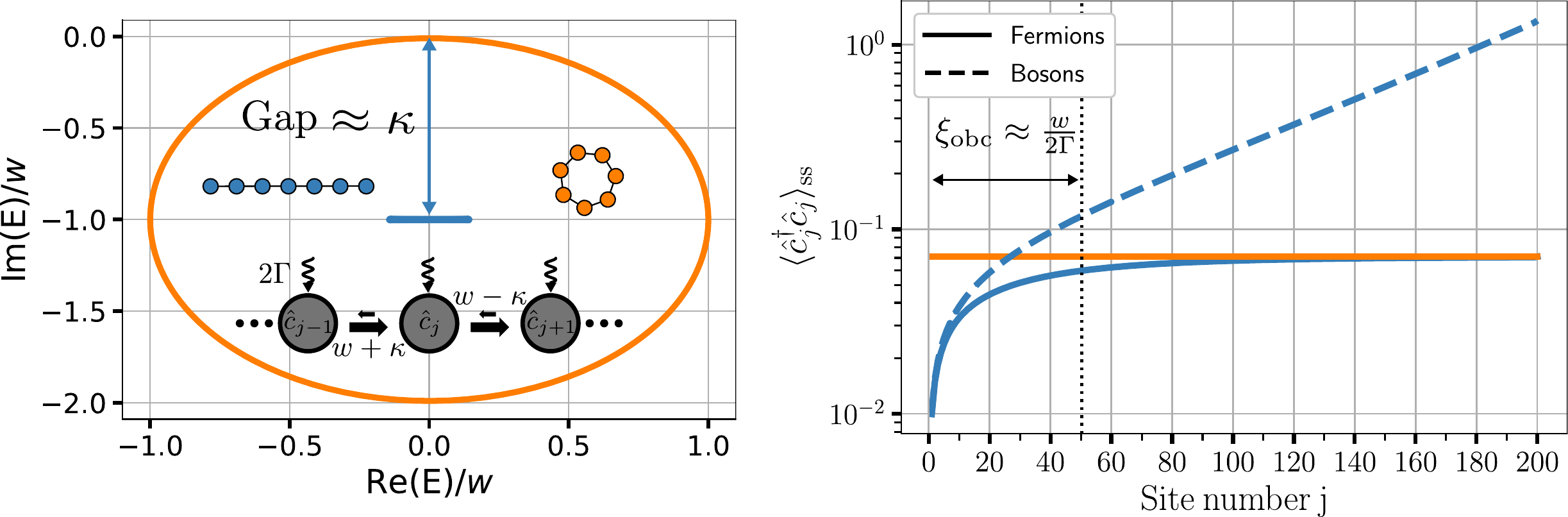}
	\caption{Left: Periodic (orange) and open (blue) chain spectrum of the quantum Hatano-Nelson model with parameters $\kappa = 0.99 w$ and $\Gamma = 0.01 w$. The spectrum for fermions and bosons only differ by shift of $-i2\Gamma$, and we thus only plot the fermionic spectrum for clarity. Dissipation is used to realize asymmetric hopping amplitudes $w \pm \kappa$, and each site is incoherently pumped at a rate $2\Gamma$.  Right: Steady-state occupation $\langle \hc_j^\dagger \hc_j \rangle_{\rm ss}$ of a quantum Hatano-Nelson model under periodic boundary conditions for fermions (orange),  and for open boundary conditions (OBC) for both fermions and bosons (blue). Remarkably,  despite the existence of a large damping gap, the density is controlled by a large length scale $\xi_{\rm obc} \approx w/(2\Gamma)$. Further, $\xi_{\rm obc}$} is unrelated to $2A = \ln[(w+\kappa)/(w-\kappa)]$, the (inverse) localization length of the non-Hermitian Hamiltonian's OBC left and right eigenvectors.  For fermions, $\xi_{\rm obc}$ corresponds to a healing length whereas for bosons it describes the exponentially localized pileup of particles on one edge. Note that we only plot the periodic boundary condition results for fermions, as the bosonic model with the same parameters is dynamically unstable. 
	\label{fig:Model_Steady_State_Occupation}
	\end{figure*}
    
    \section{Introduction}
    The physics of systems whose dynamics is governed by non-Hermitian Hamiltonians has generated interest in a wide range of fields, from classical optics \cite{Christodoulides_2017, Lan_Yang_2017, Wiersig_2014, Non_Hermitian_Optics_Dawn, Konotop_PT_RevModPhys}, to topological band theory \cite{Lieu_PRB_2018, Thomale_2019, Ueda_PRX_2018, Udea_PRX_2019, Lee_2016, Bergholtz_2020_Exceptional_Review, Ueda_2020_Review, Torres_2019_Perspective}, to soft-matter physics \cite{Scheibner_2020,Ryo_Nature, Vincenzo_NH_Band_PRL,Sone_2020,You_2020,Saha_2020,Yamauchi_2020,Tang_2020,Corentin_NAS_2020}. 
    Introducing non-Hermiticity often means forgoing seemingly basic intuition formed when studying Hermitian models. The extreme sensitivity to small perturbations \cite{Alexander_Nat_Comm, Budich_Sensor_2020, Kero_Nat_Comm} and having to revisit the bulk-boundary correspondence \cite{Xiong_2017, Kunst_2018, Herviou_NH_SVD_2019, Corentin_NAS_2020} are a few  of many such examples. Understanding these effects often amounts to studying the eigenvalues and corresponding right and left eigenvectors of an effective non-Hermitian Hamiltonian.

    In the quantum regime, a fully consistent description of non-Hermitian dynamics requires one to consider an open quantum system, where modes of interest are coupled to dissipative Markovian baths. Conditioned on the absence of a quantum jump, the dynamics are governed by a non-Hermitian Hamiltonian whose anti-Hermitian part is determined by the coupling to the environment \cite{Plenio_RMP}.  In contrast, the full unconditional dynamics also depends on fluctuations.  
    Despite numerous works examining non-Hermitian tight-binding models in quantum settings, few studies have fully addressed the unconditioned steady-state properties (for the fermionic case, see e.g.~\cite{Atland_Diehl_Symmetry_2020, Sato_Open_Quantum_2020}). 
    There thus remain several basic open questions.  These include the role of particle statistics, the possible sensitivity of the steady state to boundary conditions (analogous to the non-Hermitian skin effect (NHSE)\cite{Zhong_PRL_2018_1, Zhong_PRL_2018_2, Torres_2018_PRB}), and the general connections between the steady state's spatial structure and the underlying non-Hermitian Hamiltonian. 

    
    Here we address these questions by studying the steady states of open quantum systems that realize the physics of a target non-Hermitian tight-binding Hamiltonian $\hH_{\rm eff}$. We begin by discussing the general class of master equations that correspond to the desired $\hH_{\rm eff}$. We then construct the formal steady-state density matrix  $\hrho_{\rm ss}$ of such models, emphasizing that this requires specifying both $\hH_{\rm eff}$ and the unavoidable fluctuations arising from the coupling to dissipation.  We find generically that these steady states exhibit features and spatial structures that are not at all obvious if one simply looks at the eigenvectors of 
    $\hH_{\rm eff}$. Taking the paradigmatic non-reciprocal Hatano-Nelson model \cite{Hatano_Nelson, Hatano_Nelson_2} as an example, we find that the real-space steady-state occupation $\langle \hc_j^\dagger \hc_j \rangle_{\rm ss}$ under open boundary conditions is controlled by a new macroscopic length scale $\xi_{\rm obc}$, which is independent of the localization length of the right and left eigenvectors of $\hat{H}_{\rm eff}$. Surprisingly, this new long length scale is
    not associated with or the result of a corresponding long time-scale:  the dissipative gap is large under open boundary conditions (see Fig.~\ref{fig:Model_Steady_State_Occupation}). We argue that this feature is not specific to the Hatano-Nelson model by studying two additional models in App.~\ref{app:Two_More_Models} which feature multiple bands and/or broken time-reversal. Further, we demonstrate that the occupation is strikingly different for fermions and bosons, despite the left and right eigenvectors of $\hH_{\rm eff}$ being independent of particle statistics. Finally, we demonstrate that the set of orthogonal modes which fully specify the steady state under open boundary conditions are very similar to delocalized standing-wave states, even for extreme non-reciprocity: there is thus no true analogue of the non-Hermitian skin effect for $\hrho_{\rm ss}$.
    Our results provide a framework to understand steady states of general non-Hermitian systems in the quantum regime, and illustrate how fluctuations play a critical role.
    
    
    \section{Consistent open-system description of quantum non-Hermitian Hamiltonians } 
    \subsection{Effective unconditional non-Hermitian Hamiltonians}
     The motivating question throughout this work is straightforward: if one wants quantum dynamics of a fermionic or bosonic system generated by a target non-Hermitian Hamiltonian 
     \begin{align}\label{eq:H_targ}
         \hat{H}_{\rm targ}
         =
         \sum_{n,m}
         (H_{\rm targ})_{nm}
         \hat{c}_n^\dag \hat{c}_m,
     \end{align}
     what can be said about the steady state? Here $\hc_m$ and $\hc_n^\dagger$ are fermionic or bosonic creation and annihilation operators satisfying canonical anti-commutation and commutation relations respectively. The indices $n$ and $m$ label independent orthogonal modes, and include any and all degrees of freedom such as position, spin or polarization.

     Before attempting to formulate an answer, it is imperative to discuss how $\hat{H}_{\rm targ}$ is realized. Implementing an effective non-Hermitian Hamiltonian in a quantum system can be achieved by using parametric-amplifier type interactions \cite{Yuxin_2019, Alexander_PRX} or by considering an open quantum system (see e.g.~\cite{Kero_Nat_Comm}); 
      we focus here on the latter. 
      To ensure that the dynamics are Markovian, as is already implied by Eq.~(\ref{eq:H_targ}), the system of interest is coupled to independent Markovian reservoirs which either incoherently add or remove particles,  processes described by the jump operators $\hat{L}_{\mu} = \sum_m l_{\mu m} \hat{c}_m$ and $\hat{G}_{\nu} = \sum_{n} g^*_{\nu n} \hc_n^\dagger$ respectively. The coeffecients $l_{\mu n}$ and $g_{\nu n}^*$ specify in what state the particles are removed or added to the system by the environment, with $\mu$ and $\nu$ indexing the independent loss and pump baths.  The equation of motion for the density matrix has the standard Lindblad form, and can be written as ($\hbar = 1$)
    \begin{align} \nonumber
        i\partial_t \hrho 
        \equiv
        \mathcal{L}
        \hrho
        =
        &\left(
        \hat{\mathcal{H}}_{\rm cond} \hrho
        -
        \hrho 
        \hat{\mathcal{H}}_{\rm cond}^\dagger
        \right)
        \\ \label{eq:Linblad_Hcond}
        &+
        i \sum_{\gamma}L_\gamma \hat{l}_\gamma\hrho \hat{l}^\dagger_\gamma
        +
        i \sum_{\delta} G_\delta \hat{g}_\delta^\dagger \hrho \hat{g}_\delta
    \end{align}
    where $\mathcal{L}$ is the Lindbladian superoperator. The conditional Hamiltonian is defined as
    \begin{align}\nonumber
        \hat{\mathcal{H}}_{\rm cond}
        &=
        \sum_{n,m}
        H_{nm} 
        \hc_{n}^\dagger\hc_m
        \\ \label{eq:Second_Quantized_Heff}
        &
        -\frac{i}{2} 
        \sum_{\gamma}
        L_\gamma
        \hat{l}^\dagger_\gamma \hat{l}_\gamma
        -\frac{i}{2}
        \sum_{\delta}
        G_{\delta}
        \left(
        1\mp
        \hat{g}^\dagger_\delta \hat{g}_\delta
        \right)
    \end{align}
    with $-$ and $+$ corresponding to fermionic and bosonic creation and annihilation operators respectively. We have defined $\hat{l}_\gamma = \sum_m \braket{l_\gamma}{m} \hc_m $ and $\hat{g}_{\delta}^\dagger = \sum _n \braket{n}{g_\delta} \hc_n^\dagger$, where $\ket{l_\gamma}$ and $\ket{g_\delta}$ are eigenvectors of the Hermitian positive semi-definite matrices $L_{nm} = (\boldsymbol{l}^\dagger \boldsymbol{l})_{nm}$ and $G_{nm} = (\boldsymbol{g}^\dagger \boldsymbol{g})_{nm}$ with corresponding eigenvalues $L_\gamma$ and $G_{\delta}$. Thus, the Hermitian matrix $\boldsymbol{H}$ describes the coherent Hamiltonian of the isolated system, whereas the matrices $\boldsymbol{L}$  and $\boldsymbol{G}$ completely capture the effects of the dissipative baths. Note that the choice of coherent Hamiltonian and dissipators leads to a master equation with a $U(1)$ symmetry $\hc_n^\dagger \to e^{i \varphi} \hc_n^\dagger$, $\hc_m \to e^{-i \varphi} \hc_m$. Our results and the phenomena we discuss do not fundamentally rely on this symmetry. They can readily be extended to quadratic Hamiltonians $\hH$ which do not preserve particle number and arbitrary dissipiators which are linear in creation and annihilation operators. 
    
   By unraveling the master equation to a stochastic Schr\"{o}dinger equation, one can show that $\hat{\mathcal{H}}_{\rm cond}$ generates time evolution of the system conditioned on the absence of a quantum jump \cite{Carmichael_Book, Gardiner_Zoller}. The conditional Hamiltonian is thus only directly accessible by post-selecting measurement outcomes.  While this is feasible in some platforms \cite{Murch_2019}, it is generally challenging and involves discarding a large volume of data. More conventional experiments do not post-select, and thus probe the full unconditioned dynamics of $\hrho$. 
   Yet in this setting, it is not obvious that it is even possible to attribute the evolution of the density matrix to a single-particle Hamiltonian as we did for the post-selected evolution.  
   How then should we think about the unconditioned evolution of $\hrho$?
   
  The answer lies in the unconditional equations of motion of the normal-ordered covariance matrix $\langle \hc_n^\dagger \hc_m \rangle$ which reads (see App.~\ref{app:Normal_Order_Corr})
    \begin{align} \nonumber
        i\partial_t \langle \hc_n^\dagger \hc_m \rangle
        &=
        \sum_{a}
        \left(
        (H_{\rm eff})_{ma} \langle \hc_n^\dagger \hc_a \rangle
        -
        (H_{\rm eff}^\dagger)_{an}\langle \hc_a^\dagger \hc_m \rangle
        \right)
        \\ \label{eq:EOM_Correlation}
        &+ i G_{mn}
    \end{align}
    where 
     \begin{align} \label{eq:Heff_Matrix_Def}
        \Heff
        \equiv
        \boldsymbol{H}
        -\frac{i}{2}
        \left(
        \boldsymbol{L} \pm \boldsymbol{G}
        \right).
    \end{align}
    The $+$ and $-$ is for fermions and bosons respectively. This effective Hamiltonian $\Heff$, which fully incorporates the effects of the jumps, is in general distinct from the effective Hamiltonian relevant to no-jump conditional evolution.  From Eq.~(\ref{eq:Second_Quantized_Heff}), this conditional Hamiltonian is 
    \begin{align}
    \boldsymbol{H}_{\rm cond}
    =
    \boldsymbol{H}
    -\frac{i}{2}
    \left(
    \boldsymbol{L}
    \mp
    \boldsymbol{G}
    \right).
    \end{align}
    Note that $\Heff$ and $\boldsymbol{H}_{\rm cond}$ only coincide in the absence of incoherent pumping (i.e.~$\boldsymbol{G}=0$).  
       We stress that the identification of $\Heff$ from the covariance matrix dynamics is independent of how correlators are ordered.  Other ordering prescriptions lead to the same dynamical matrix $\Heff$ (only the inhomogeneous term in the equations is modified, see App.~\ref{app:Normal_Order_Corr}). 
    As shown in Refs.~\cite{Prosen_Fermions_2008, Prosen_Bosons_2010}, solving Eq.~(\ref{eq:EOM_Correlation}) is tantamount to knowing the full structure of the Lindbladian $\mathcal{L}$.
    In particular, the steady state density matrix is Gaussian, and hence fully characterized by two-point averages.  Eq.~(\ref{eq:EOM_Correlation})  
    thus lets us unambiguously identify the non-Hermitian dynamical matrix relevant to the unconditional steady state as $\Heff$. 

    Eq.~(\ref{eq:EOM_Correlation}) is particularly easy to interpret when our particles are bosons.  In this case, our system could also be described using the Heisenberg-Langevin equations
    \begin{align}
        i \partial_t \hc_m
        =
        \sum_a
        (H_{\rm eff})_{ma}
        \hc_l
        -
        \sum_{\mu}
        l_{\mu m}
        \hat{\eta}_{\mu}
        -
        \sum_{\nu}
        g^{*}_{\nu m}
        \hat{\zeta}^\dagger_\nu
    \end{align}
    which are equivalent to the master equation. The inhomogeneous terms $\hat{\eta}_\mu$ and $\hat{\zeta}_\nu^\dagger$ are the operator equivalent of independent Gaussian white noise with zero mean and unit variance (see Ref.~\cite{RMP_Clerk} for a pedagogical introduction).
   The upshot is that Eq.~(\ref{eq:EOM_Correlation}) affords a similar interpretation for \textit{either} particle type: $\hc_m$ evolves under $\Heff$ while being driven by white noise, 
   and particle statistics only play a role in determining the anti-Hermitian part of $\Heff$, 
   the difference being a simple change of sign.
   For bosons, pump baths tend to generate amplification and exponential growth of amplitude and particle number.  In contrast, for fermions, the Pauli principle makes this impossible and precludes exponential growth.  This is enforced by the simple sign change in Eq.~(\ref{eq:Heff_Matrix_Def}).
 

\subsection{Constructing valid quantum descriptions of a target non-Hermitian Hamiltonian }
   
   Eq.~(\ref{eq:EOM_Correlation}) shows how a non-Hermitian Hamiltonian 
   $\Heff$ naturally arises in the description of unconditional dissipative quantum dynamics.  We now ask a reverse engineering question:  if one {\it starts} with a given, non-Hermitian Hamiltonian of interest
   $\boldsymbol{H}_{\rm targ}$, how does one construct a valid corresponding quantum open system?  This amounts to making consistent choices of both $\Heff$ and the pumping matrix 
   $\boldsymbol{G}$ appearing in Eq.~(\ref{eq:EOM_Correlation}) to match the desired dynamics.  This construction is a crucial first step in understanding how any interesting features of $\boldsymbol{H}_{\rm targ}$ might manifest themselves in a quantum setting (including the steady state).  

    Naively, one might start by simply picking $\Heff$ in
    Eq.~(\ref{eq:EOM_Correlation}) to be identical to the desired Hamiltonian $\boldsymbol{H}_{\rm targ}$.  The validity of this procedure surprisingly depends on particle statistics. For fermions, we need to mindful of a constraint arising from the exclusion principle:  $\Heff$ for fermions cannot give rise to exponential growth, implying that the anti-Hermitian part of $\Heff$ must be negative semi-definite.  This follows directly from Eq.~\eqref{eq:Heff_Matrix_Def}.  Thus, for fermions, we will in general have to add loss to 
    $\boldsymbol{H}_{\rm targ}$ to satisfy this constraint. In contrast, choosing $\Heff = \boldsymbol{H}_{\rm targ}$ for bosons is always permissible. 
  
    We in general however will require more than a valid master equation, but will also want to ensure the existence of a unique steady-state.  For this, the eigenvalues of the dynamical matrix $\Heff$ must have negative-definite imaginary parts.  To achieve this, we will generically have to add extra loss to bosonic Hamiltonians as well. To that end, let $\lambda$ denote the largest positive eigenvalue of $ -i(\boldsymbol{H}_{\rm targ}-\boldsymbol{H}_{\rm targ}^\dagger)/2$.  We will then choose $\Heff$ according to:
     \begin{align}    \label{eq:HeffShifted}
        \Heff \equiv 
        \lim_{\nu \to 0^+}
        \left(
         \boldsymbol{H}_{\rm targ}- i (\lambda+\nu) \boldsymbol{1} 
         \right).
    \end{align}
    This is a minimal prescription for satisfying the constraint for fermions and bosons: we simply add enough uniform loss to each mode to ensure the no-gain condition is satisfied.  It implies that the eigenvectors of $\Heff$ coincide that of $\boldsymbol{H}_{\rm targ}$, and their spectra differ at most by a trivial global shift. This ensures that  any interesting and desirable novel non-Hermitian phenomena exhibited by $\boldsymbol{H}_{\rm targ}$ will also be present in $\Heff$.  Note that having $\nu = 0^+$ ensures a unique steady state.
    
    We stress that the no-gain condition on the anti-Hermitian part of $\Heff$ is strictly speaking distinct from requiring that the eigenvalues of $\Heff$ have negative-definite imaginary part. The former, for fermions, ensures consistency with a valid master equation while the later, for both types of particles, is required for the existence of the steady-state.

    Having $\Heff$ match $\boldsymbol{H}_{\rm targ}$ means that the drift terms in Eq.~(\ref{eq:EOM_Correlation}) directly mirror the desired non-Hermitian dynamics.  This does not however specify our open quantum system:  we also need to specify $\boldsymbol{G}$ (i.e.~the noise) in a manner that is consistent with $\Heff$. There are of course many different ways to achieve this.  In what follows, we present two simple and physically motivated approaches.  

     
     
   \subsubsection{Method 1 - Minimal Prescription}
    
    Recall that the anti-Hermitian part of $\bm H_{\rm eff}$ is 
    determined by  Eq.~\eqref{eq:Heff_Matrix_Def}. 
    The simplest way to fully specify our open quantum system is to have both the loss matrix   
    $\bm L$ and pumping matrix $\bm G$ be proportional to the anti-Hermitian part of $\bm H_{\rm eff}$.  This leads to:
    \begin{align}
    \boldsymbol{L} &\equiv
    \begin{cases}
    i(1-\epsilon)(\Heff-\Heff^\dagger) 
    &\mbox{fermions}
    \\
    i \epsilon(\Heff-\Heff^\dagger)
    & \mbox{bosons }
    \end{cases}
    \end{align}
    \begin{align}
    \boldsymbol{G} &\equiv 
     \begin{cases}
    i \epsilon(\Heff-\Heff^\dagger) 
    &\mbox{fermions}
    \\
    i(\epsilon-1)(\Heff-\Heff^\dagger)
    & \mbox{bosons }
    \end{cases}
    \end{align}
    where the parameter $\epsilon$ ($0 \leq \epsilon \leq 1$ for fermions and $\epsilon > 1 $ for bosons) determines the balance between pumping and loss.  Diagonalizing $\boldsymbol{L}$ or $\boldsymbol{G}$ and specifying $\epsilon$ then directly determines one possible set of loss and pumping jump operators $\hat{L}_\mu$ and $\hat{G}_\nu$.
    This method gives us a simple way of generating a valid open system (for either fermions or bosons) with a minimal number of extra assumptions.  The only additional parameter introduced (beyond the desired target Hamiltonian $\boldsymbol{H}_{\rm targ}$) is $\epsilon$.  It controls the average particle number in the system, and hence plays the rough role of a chemical potential.

    \subsubsection{Method 2 - Featureless Pumping}
    \label{subsubsec:Method2}

\begin{figure}[t]
	\centering
	\includegraphics[width=0.475\textwidth]{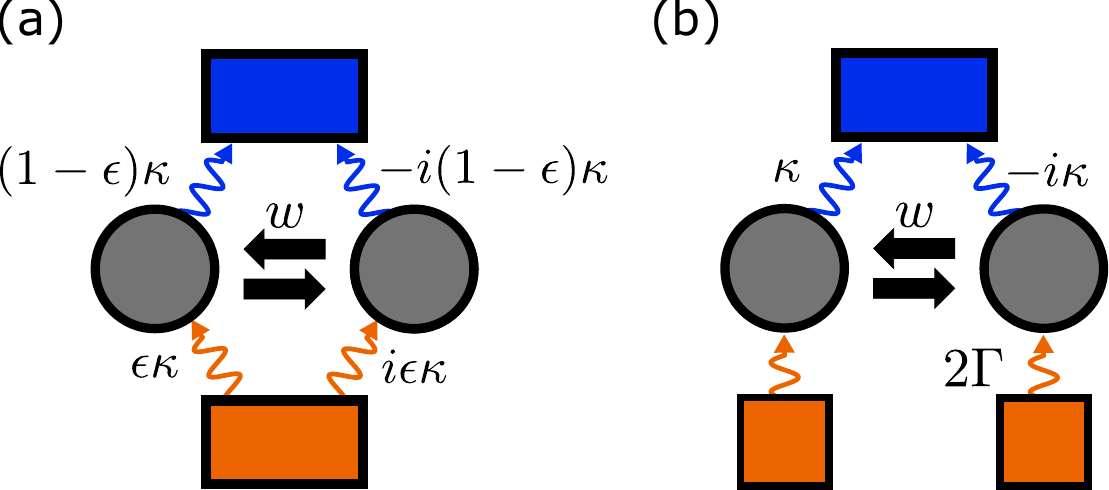}
	\caption{
	Two different ways to realize the dissipative version of the fermionic quantum Hatano-Nelson. Although both methods lead to an effective Hatano-Nelson Hamiltonian, they do not have the same steady state, since the noise $\boldsymbol{G}$ is not equivalent. (a) In the first method, each loss bath (blue) has an equivalent pumping bath (orange). The parameter $\epsilon$ controls the strength of the fluctuations $\boldsymbol{G}$. (b) In method two, the non-reciprocal hopping is realized using loss dissapators only. Each mode is also subject to uniform pumping, which leads to a non-trivial steady state. 
	}
	\label{fig:HN_Realization}
	\end{figure}

    An alternate approach that in many cases is more experimentally tractable is to only use structured loss to realize the anti-Hermitian part of $\Heff$, i.e.~$\bm L = i(\Heff-\Heff^\dagger)$.  Of course, one still needs some pumping to have a steady state with non-zero particle number.  This can be achieved by also introducing spatially uniform, featureless pumping to the system:  $\boldsymbol{G} = 2 \Gamma \boldsymbol{1}$, where $\Gamma$ is an overall pumping rate.  This pumping only changes $\Heff$ by a constant diagonal term, implying that it still faithfully reflects the dynamics of the desired non-Hermitian Hamiltonian $\boldsymbol{H}_{\rm targ}$. This is the method we employ throughout the main text. A visual comparison of the two methods for fermions is shown in Fig~\ref{fig:HN_Realization}. Note that for bosons, pumping always decreases decay rates and care must be take to ensure dynamical stability (by, e.g. adding additional background loss). 

    \subsection{Steady state $\hrho_{\rm ss}$}
    Having constructed a consistent master equation corresponding to the target dynamics $\hat{H}_{\rm targ}$, we now characterize the steady state of Eq.~(\ref{eq:Linblad_Hcond}). With a quadratic coherent Hamiltonian and linear jump operators, the stationary-state is Gaussian and completely determined by the steady-state covariance matrix 
        \begin{align}
        F_{mn} 
        \equiv
        \langle 
        \hc_n^\dagger \hc_m
        \rangle_{\rm ss}.
    \end{align}
    If we assume that all eigenvalues of $\Heff$ have a non-zero negative imaginary part (as ensured by our construction in the previous subsection) then the steady state is unique, with $\boldsymbol{F}$ satisfying the so-called Lyaponov equation
    \begin{align}\label{eq:Lyaponov}
        \Heff \boldsymbol{F}
        -
        \boldsymbol{F} \Heff^\dagger
        =
        -i \boldsymbol{G},
    \end{align}
    which follows from Eq.~(\ref{eq:EOM_Correlation}). The formal solution to $\boldsymbol{F}$ reads
    \begin{align}\label{eq:Fnm_Formal_Solution} \nonumber
        F_{mn}
        &
        =
        \int_{-\infty}^{\infty}
        \frac{d \omega}{2\pi}
        \bra{m}
        \frac{1}{\omega\boldsymbol{1}-\Heff}
        \boldsymbol{G}
        \frac{1}{\omega\boldsymbol{1}-\Heff^\dagger}
        \ket{n}
        \\
        &
         =
         \sum_{\delta}
        G_{\delta}
        \int_{-\infty}^{\infty}
        \frac{d \omega}{2\pi}
        \bra{m}
        \frac{1}{\omega \boldsymbol{1}-\Heff}
        \ket{g_\delta}
        \bra{g_\delta}
        \frac{1}{\omega \boldsymbol{1}-\Heff^\dagger}
        \ket{n}.
    \end{align}
     The last expression provides a simple intuitive interpreitation.  At each frequency, the pump baths populates the state $\ket{g_\delta}$ at a rate $G_\delta$ which then evolves under the propagator $(\omega \boldsymbol{1}-\Heff)^{-1}$ to different sites $\ket{m}$ in the lattice. 

   We stress that $\hrho_{ss}$ is thus controlled both by the non-Hermitian Hamiltonian \textit{and} the fluctuations from the loss and pumping baths. 
   This exact solution $F_{mn}$ should be contrasted with previously suggested prescriptions on how to associate steady states with a given non-Hermitian Hamiltonian.  These methods, such as exponentiating the non-Hermitian Hamiltonian \cite{Herviou_2019_Non_Hermitian_Entanglement, Ueda_Non_Hermitian_Criticality} or occupying the right or left eigenstates \cite{Ueda_Non_Hermitian_Superconductivity_2019}
   are either ad-hoc or assume conditional dynamics.  They are thus not relevant to the generic (unconditional) situation we consider.
    
    If $\Heff$ can be diagonalized, then the steady state correlation matrix can also be written as: 
    \begin{align} 
        \boldsymbol{F}
        &=
        -i
        \sum_{\alpha, \beta}
        \ket{\psi^R_\alpha}
        \left(
        \frac{\langle \psi^L_\alpha| \boldsymbol{G} |\psi^L_\beta \rangle}
        {E_\alpha-E_\beta^*}
        \right)
        \bra{\psi^R_\beta}
        \label{eq:F_Right_Left}
    \end{align}
     where $\ket{\psi^R_\alpha}$ and $\ket{\psi^L_\alpha}$ are the right and left biorthonormal $ \langle \psi^L_\alpha | \psi^R_\beta \rangle = \delta_{\alpha \beta}$ eigenvectors of $\Heff$ with eigenvalue $E_\alpha$. The form of $\boldsymbol{F}$ reminds us how we should interpret the eigenvectors of a non-Hermitian matrix. A right eigenvector $\ket{\psi^R_\alpha}$ is a mode whose temporal evolution is trivial and determined by the eigenvalue $E_\alpha$, akin to how one normally thinks of eigenstates of a Hermitian matrix. The meaning of the left eigenstate $\ket{\psi^L_\alpha}$ is more subtle: it describes the susceptibility of the corresponding right eigenvector to a spatially varying perturbation. More concretely, the overlap $\braket{\psi_\alpha^L}{g_{l}}$ quantifies how the gain bath $l$ ``populates" the state $\ket{\psi_{\alpha}^R}$ \cite{Schomerus_2020}. 
    
    Note that if we choose to implement our non-Hermitian dynamics using structured loss and uniform pumping (i.e.~$\boldsymbol{G} = 2 \Gamma \boldsymbol{1}$. c.f.~Sec.~\ref{subsubsec:Method2}), then Eq.~(\ref{eq:F_Right_Left}) reduces to:
     \begin{align} \label{eq:F_R_L_Eigenvectors}
        \boldsymbol{F} = 
        -2i \Gamma
             \sum_{\alpha, \beta}
        \ket{\psi^R_\alpha}
        \left(
        \frac{\langle \psi^L_\alpha|\psi^L_\beta \rangle}
        {E_\alpha-E_\beta^*}
        \right)
        \bra{\psi^R_\beta}.
    \end{align}
  Even in this seemingly simple case, we see that the steady state does correspond to a simple statistical mixture of right eigenvectors.

   We have thus in principle achieved our goal of identifying the steady state: it is completely determined by the non-Hermitian Hamiltonian $\Heff$ and the noise matrix $\boldsymbol{G}$ through Eqs.~(\ref{eq:Fnm_Formal_Solution}) and (\ref{eq:F_Right_Left}). 
   As we will now show, these formal expressions do not immediately provide useful intuition.  In particular, the non-trivial interplay between the dynamics and the noise can lead to a steady state which cannot be understood by considering either independently.  
   Further, there exist subtle new length scales in the steady state that are not associated with any one eigenvector, and which are not obvious from the general expression in Eq.~(\ref{eq:Fnm_Formal_Solution}). 
  
   \subsection{Relevance to other work}

    We pause to note connections and differences between our work and previous studies connecting non-Hermitian physics, open quantum systems and steady states. While most works focus on conditional dynamics and the conditional Hamiltonian $\hat{\mathcal{H}}_{\rm cond}$, several papers have addressed aspects of unconditional evolution and the role played by $\Heff$. We stress that our work is markedly distinct from these previous studies. Refs.~\cite{Lieu_PRL_2020,Sato_Open_Quantum_2020, Atland_Diehl_Symmetry_2020} discuss $\hrho_{\rm ss}$ in the context of topological classification, which is not the focus here.  The steady states of continuum interacting systems exhibiting non-Hermitian structures has also been studied in certain cases (see e.g.~\cite{Hanai_2019,Hanai_2020}); this is also distinct from our non-Hermitian band-structure setting.  
    Other works such as Refs.~\cite{Longhi_Skin_Effect_2020, Zhong_Skin_Effect_OQS_2019,Nori_Liouvillian_PRA} explore the relaxation dynamics towards the stationary state, as determined solely by $\Heff$. In contrast, our focus is the steady state itself, which manifestly depends on \textit{both} the dynamics $\Heff$ and the noise $\boldsymbol{G}$. The analysis in Ref.~\cite{Prosen_Noise_Driven_2010} does include the effect of fluctuations but focuses solely on infinite translationally invariant systems. Here, we show that the steady state is drastically different with boundaries present, in a manner that is not a trivial consequence of the NHSE.
    
    Further, we analyze non-Hermitian Hamiltonian of both fermions and bosons, whereas  previous works have mostly focused exclusively on the fermionic case.
   
   We also note that quantum non-reciprocal models have been previously studied, motivated by applications to quantum engineering.  Most works focus on few-mode systems, but lattices have also been considered recently \cite{Metelmann_PRA_2018, Alexander_Nat_Comm}.
   Unlike our work, the motivation in these previous studies is different, and the focus is primarily on the output state of radiation emitted from the lattice.  
   

    
    \section{Quantum Hatano-Nelson Model}
        We now focus on determining the steady state of an $N$-site Hatano-Nelson model, which describes particles asymmetrically hopping on a 1D lattice \cite{Hatano_Nelson, Hatano_Nelson_2}. In addition to being among the simplest of non-Hermitian tight-binding models, it also displays rich features such as the non-Hermitian skin effect \cite{Zhong_PRL_2018_1, Kunst_2018, Alexander_PRX} and non-Hermitian topology \cite{Ueda_2020_Review}. It thus serves as the ideal candidate to study how the non-Hermitian Hamiltonian imprints itself on the steady state. 
        
        The target Hamiltonian of interest reads
        \begin{align} \label{eq:HN_First_Quantized}
            \hat{H}_{\rm targ}^{\rm HN}
            =
            \sum_{j}
            \bigg[
            \frac{w+\kappa}{2} \hc_{j+1}^\dagger \hc_j
            +
            \frac{w-\kappa}{2}\hc_j^\dagger \hc_{j+1}
            \bigg]
        \end{align}
        where  without loss of generality we take $w, \kappa >0$. Note the  anti-Hermitian part of this Hamiltonian (proportional to $\kappa$) is not negative semi-definite, regardless of boundary conditions.  To permit a quantum treatment valid for both fermions and bosons, we thus modify the Hamiltonian by adding minimal uniform loss, resulting in a valid $\Heff$
        whose anti-Hermitian part is now negative semi-definite as required
        (c.f.~Eq.~(\ref{eq:HeffShifted})).  Further, we will use method 2 (c.f.~Sec.~\ref{subsubsec:Method2}) to realize this non-Hermitian, non-reciprocal Hamiltonian using a set of structured, non-local loss baths and uniform incoherent pumping. 
        
        Following the above prescription, 
        we obtain a Lindblad master equation (c.f.~Eq.~(\ref{eq:Linblad_Hcond})) corresponding to the target Hatano-Nelson Hamiltonian.  The coherent Hamiltonian $\hH$ and jump operators $\hat{L}_j$ and $\hat{G}_j$ are given by (see App.~\ref{app:HN_Method_2})
     \begin{align} \label{eq:H_coh_HN}
			\hH 
			&= 
			\frac{w}{2}
			\sum_{j}
			\left(
			\hc_{j+1}^\dagger\hc_j+h.c.  
			\right)
			\\ \label{eq:L_n_HN}
			\hat{L}_j &= 
			\sqrt{\kappa} \left(\hc_j-i \hc_{j+1}\right)
			\\ \label{eq:G_n_HN}
			\hat{G}_j &= 
			\sqrt{2\Gamma} \hc^\dagger_j
	\end{align}
where $w$ is the coherent nearest-neighbour hopping, $\kappa$ is the decay rate, $\Gamma$ is the pumping rate. The resulting non-Hermitian Hamiltonian is (up to a constant) that of the Hatano-Nelson model: $\Heff = \boldsymbol{H}_{\rm targ}^{\rm HN} -i(\kappa \pm \Gamma) \boldsymbol{1} $.  
Note that similar structured loss dissipators have been used to study non-reciprocal hopping in previous works, albeit with very different motivations (see e.g.~\cite{Metelmann_PRX_2015,Nunnenkamp_2018,Metelmann_PRA_2018,Fazio_PRA_2018}).

While our mapping is general, we focus in what follows on the interesting case of strong non-reciprocity ($\kappa \lesssim w$) and weak pumping ($\Gamma \ll w, \kappa$). 
In this limit, the natural expectation is that the pumping serves only as a weak probe of the underlying non-Hermitian dynamics: its only purpose is to populate the system without disrupting the dynamics.  As we will show, this is surprisingly not the case:  the weak pumping rate $\Gamma$ determines a new, extremely long system length scale.  Further, for open boundary conditions, this long length scale is not related to a corresponding long relaxation time scale (i.e.~the dissipative gap remains constant as $\Gamma \rightarrow 0$).


\subsection{Periodic boundary conditions}
We first work with periodic boundary conditions. With the effective Hamiltonian $\Heff$ and pumping matrix $\boldsymbol{G}$, we can readily use the formalism described in the previous section to determine the steady-state correlation matrix $\boldsymbol{F}$.  It is perhaps however more instructive to write the master equation in momentum-space.  Letting $N$ denote the number of sites in our lattice and $\hc_k = \sum_n e^{-i k n } \hc_n/\sqrt{N}$, we have:
	\begin{align} \nonumber
	i \partial_t \hrho
    		&=
			\sum_{k}
			w \cos k[ \hc^\dagger_k \hc_k, \hrho \ ]
			\\  \label{eq:EOM_Momentum}
			&
			+
			i 
			\sum_{k}
			\kappa(k)
			\mathcal{D}[\hc_k] \hrho 
			+
			i
			2\Gamma
			\sum_{k}
			\mathcal{D}[\hc_k^\dagger] \hrho
	\end{align} 
    where $\kappa(k) = 2\kappa(1+\sin k)$ and the allowed momenta are $k = 2\pi q/N$ with $q$ an integer running from $1$ to $N$. This form of the master equation serves as a useful reminder that non-Hermitian hopping can also be interpreted as momentum-depending damping $\kappa(k)$. This is also apparent by considering the spectrum of $\Heff$ which reads $E(k) = w \cos k-i \kappa(1+\sin k)\mp i \Gamma$. To ensure stability in the bosonic case $\Im(E(k)) < 0$, we always work in the limit of weak pumping and where the quantization of $k$ leads to $\Gamma < \kappa(1+ \sin k)$ for all allowed $k$.
    
    It is clear from Eq.~(\ref{eq:EOM_Momentum}) that there will be no steady-state coherences between different momentum states. The independent baths add and remove particles from these orthogonal modes, and the coherent dynamics can not cause any transition between such states. The only non-vanishing steady-state component in this basis are the diagonal elements 
    \begin{align}\label{eq:Momentum_Occupation_SS}
        \langle \hc_k^\dagger \hc_k \rangle_{\rm ss}
        =
        \frac{1}{e^{\beta \epsilon(k)} \pm 1}
    \end{align}
    where we have defined the effective Boltzman factor
    \begin{align} \label{eq:Boltzmann}
        e^{-\beta \epsilon(k)}
        \equiv
        \frac{2\Gamma}{\kappa(k)}.
    \end{align}
    The steady state of the periodic-boundary condition Hatano-Nelson model is thus completely determined by these effective momentum-dependent Boltzmann factors.  The non-reciprocity of our system reflects itself in an asymmetry of the momentum-space occupancy of mode at $k$ versus $2 \pi -k$.  The real-space steady-state density is of course uniform, as required by translational invariance.
    We also stress that only the product of $\beta$ and $\epsilon(k)$ is physically meaningful.
    
    A simple but crucial point is that our steady state is insensitive to the coherent Hamiltonian $\hat{H}$; the steady state would not change even if we added additional (translationally-invariant) terms to $\hat{H}$.  This is a general feature of master equations where the Hamiltonian $\boldsymbol{H}$, loss $\boldsymbol{L}$ and pumping matrices $\boldsymbol{G}$ commute with one another. In this instance, it is always possible to write the master equation in a compact form by using the basis which diagonalizes all three matrices, in analogy with Eq.~(\ref{eq:EOM_Momentum}). The physics underling the steady states of these models are then readily understood in terms of a orthogonal set of modes, the same as the effective Hamiltonian $\Heff$.
    

For instance, in this translationally-invariant model, the real-space correlation functions for fermions can be easily obtained using Eq.~(\ref{eq:Momentum_Occupation_SS}):
\begin{align}\label{eq:Avg_Particle_PBC_2}
    \langle \hc^\dagger_j \hc_p \rangle_{\rm ss}
    =
    \frac{1}{N}
    \sum_{k}
    \frac{\Gamma e^{-i(j-p) k}}{\Gamma+\kappa(1+\sin k)}
    \sim 
    e^{-|j-p|/\xi_{\rm pbc}}
\end{align}
where the difference $j-p$ is understood to be modulo $N$. Note that bosons are generically unstable in this model due to the additional pumping $\Gamma$, and there is thus no steady state. Here $\xi_{\rm pbc}$ is length scale determined by the dissipation, and is defined by $\kappa \cosh \xi_{\rm pbc}^{-1} \equiv \kappa+\Gamma$.  For small pumping it behaves as $\xi_{\rm pbc} = \sqrt{\kappa/(2\Gamma)}$ (see App.~\ref{app:N_ss} for details). This (inverse) length scale can be readily extracted by considering the spectrum of our periodic chain $E(k) = w \cos k - i \kappa(1+\sin k)-i \Gamma$. The smallest decay rate of any mode is $\Gamma$. It is thus not surprising that this quantity gives rise to a large length scale. In fact, it was shown in Ref.~\cite{Prosen_Noise_Driven_2010} that for a transitionally- invariant non-Hermitian tight-binding model with periodic boundary conditions, this is \textit{always} the case: a divergent correlation length must be accompanied by a critical slowing down, i.e. a vanisingly small decay rate. Surprisingly, we will see that this intuitive connection between large length and times scales will be violated by introducing boundaries.

		\subsection{Open boundary conditions}

	   We now examine the properties of our 
	   quantum Hatano-Nelson chain under open (rather than periodic) boundary conditions.  It is well known that this change of boundary conditions dramatically alters the spectrum and eigenvectors of $\Heff$ \cite{Hatano_Nelson, Hatano_Nelson_2}. 
	   At a formal level, this can be understood as arising from an incompatibility between the Hermitian and anti-Hermitian parts of $\Heff$.  For periodic boundary conditions, both are diagonalized by plane waves, whereas with open boundary conditions these matrices no longer commute.  This incompatibility has a direct consequence on our quantum master equation 
	   (specified by Eqs.~(\ref{eq:H_coh_HN})-(\ref{eq:G_n_HN})):  
	   the jump operators associated with our structured loss can now cause transitions between different eigenstates of the coherent Hamiltonian $\hat{H}$.  As we will see, this directly leads to a far more interesting steady state than in the periodic boundary condition case. 
	   
\begin{figure}[t]
		\centering
		\includegraphics [width=0.475\textwidth]{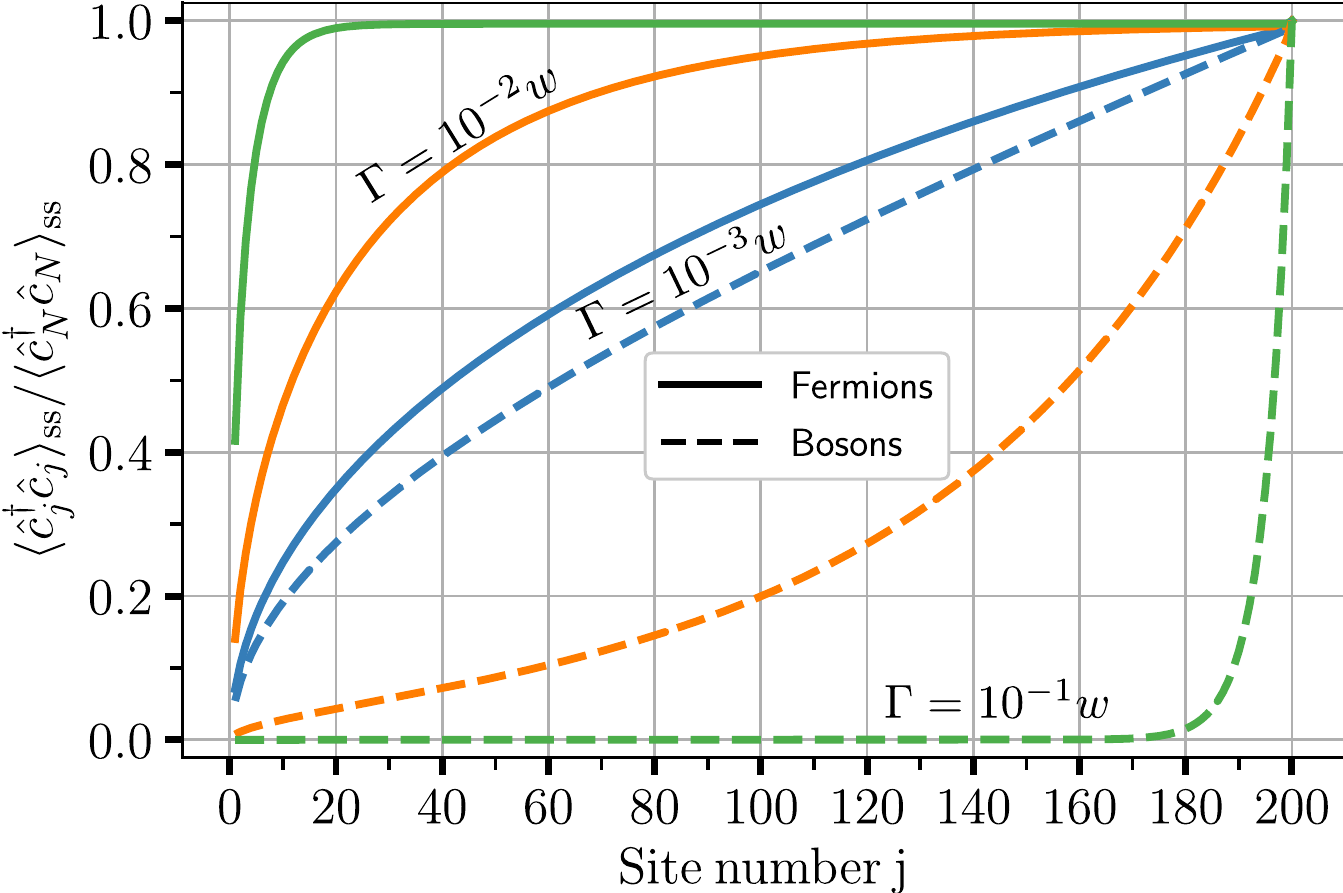}
		\caption{
            Scaled steady-state occupation $\langle \hc_j^\dagger \hc_j \rangle_{\rm ss}/\langle \hc_N^\dagger \hc_N \rangle_{\rm ss} $ under open boundary conditions of the Hatano-Nelson model realized with a coherent Hamiltonian and dissipators Eqs.~(\ref{eq:H_coh_HN})-(\ref{eq:G_n_HN}) for $N = 200$ sites and $\kappa = 0.99 w$. For small enough pumping, the occupation for bosons and fermions are essentially identical. As we increase the pumping, the length scale $\xi_{\rm obc}$ decreases. In the fermionic case this leads to a depletion of particles on the left-hand side of the chain. For bosons, this leads to exponential localization of particles on the right hand side of the system. Note that unlike the periodic boundary condition case, the bosonic model is dynamically stable for these parameters ($\kappa \leq w$ and $\Gamma < \kappa$). These numeric plots are indinguishable from the analytical forms predicted by Eqs.~(\ref{eq:Steady_State_Occupation_Full_Gamma}) on this scale.}
		\label{fig:Steady-State_Occupation_Vary_Gamma}
	    \end{figure}

    The most immediate consequence of having boundaries is that the steady-state average density  $\langle \hc_{j}^\dagger \hc_{j} \rangle_{\rm ss}$ will not be spatially uniform, as shown in Fig.~\ref{fig:Steady-State_Occupation_Vary_Gamma}. It is tempting to assume that this will 
    simply reflect the NHSE; 
    however, we show in the following that this is due to a distinct mechanism.

    In the conventional scenario, the NHSE
    causes right eigenvectors to localize at one edge:  $\braket{j}{\psi^R_\alpha} \propto e^{A j}$ with the inverse localization length $A$ given by
 	   \begin{align}\label{eq:Def_A}
	       A = \frac{1}{2} \ln \left(\frac{w+\kappa}{w-\kappa} \right).
	   \end{align}

The appearance of this short lengthscale $A^{-1}$ is accompanied by an opening of the dissipative gap: the decay rate of all eigenmodes of $\Heff$ become large, unlike its periodic counterpart which has nearly undamped modes, see Fig.~(\ref{fig:Model_Steady_State_Occupation}).
Conventional wisdom would then suggest that any length scale in the open boundary case is much smaller than the periodic boundary lengthscale $\xi_{\rm pbc} \gg \xi_{\rm obc}$; particles are so heavily damped that they can not propagate very far. 
However, surprisingly, the opposite is true: not only is the characteristic length for the steady state in open boundary condition $\xi_{\rm obc} = w/(2\Gamma)$ completely unrelated to the localization length $A$ of right and left eigenvectors, but it is also parametrically larger than the lengthscale of the equivalent system with periodic boundary conditions $\xi_{\rm pbc} = \sqrt{\kappa/(2\Gamma)}$
at small $\Gamma$.  Hence, simply knowing about the NHSE and its impact on the eigenvectors or eigenvalues does not immediately let one understand the spatial dependence of the steady-state density. 
We expect this to be a generic feature of open quantum systems exhibiting NHSE: we have provided two more examples in App.~\ref{app:Two_More_Models}.

	   Instead of thinking about eigenvectors of eigenvalues of $\Heff$, a more fruitful starting point is to use the alternate formal solution for the steady-state correlation matrix given in Eq.~(\ref{eq:Fnm_Formal_Solution}).  This yields
	   	    \begin{align} \label{eq:Real_Space_Occupation}
	        \langle \hc_j^\dagger \hc_j\rangle_{\rm ss}
	        &
	        =
	       2 \Gamma 
	        \sum_{p=1}^N
	        \int_{-\infty}^{\infty}
			\frac{d \omega}{2\pi}
	        |G^{R}_{\rm obc}[j,p;\omega]|^2
	    \end{align}
	    with $G^{R}_{\rm obc}[j,p; \omega] \equiv \bra{j}(\omega \boldsymbol{1} - \Heff)^{-1} \ket{p} $ the real-space retarded Green's function with open boundaries. Equation~(\ref{eq:Real_Space_Occupation}) reminds us how the incoherent pumping populates each site. A pump bath attached to site $p$ injects a particle with frequency $\omega$ which then propagates to site $j$ with an amplitude given by $G^R_{\rm obc}[j,p;\omega]$. The total particle number is thus this amplitude squared summed over all baths and frequencies.

       The physical picture provided to us by Eq.~(\ref{eq:Real_Space_Occupation}) implies that we should attempt to understand the real-space propagation dynamics of $\Heff$ as encoded by the Green's function. Although the NHSE forces both the eigenvectors and eigenvalues of $\Heff$ to change drastically when there are edges, on physical grounds the same should not be true of the Green's function. For the short-ranged Hamiltonian under consideration, we expect for large enough system size $N$ that the response for open boundary conditions is well approximated by the infinite-system Green's function $G^R_{\rm obc}[j,p;\omega] \approx G^{R}_{\infty}[j-p;\omega]$, at least within the bulk $1 \ll j, p \ll N$. The length scale which controls the steady-state occupation $\langle \hc^\dagger_j \hc_j \rangle_{\rm ss}$ is therefore essentially determined by how far a particle injected into the boundary-free chain can propagate before it loses an appreciable amount of its amplitude. 
       
       The dispersion of the Hatano-Nelson model $E(k)$ $= w \cos k$ $ -i \kappa(1+\sin k) \mp i \Gamma $ can be readily used to extract this length scale.  Indeed, ignoring the boundaries affords us the possibility to think about propagation in terms of plane waves. 
       We first note that the least damped momentum mode at $k = -\pi/2$ also has 
       the largest 
       group velocity $w (\partial_k \cos k) |_{k = - \pi/2} = w > 0$ which implies a right-moving mode. 
       We can thus estimate the relevant decay length as this maximal group velocity $w$ divided by the residual decay rate $\Gamma$ of that mode. We thus have $G^R_{\rm obc}[j,p;0] \sim e^{\mp|j-p|/(w/\Gamma)}$ when $p<j$.  Note the sign difference for fermions versus bosons: $\Gamma$ induces additional damping for fermions yet bosons experience spatial amplification as a consequence of the additional pumping.
       
       
       The above heuristic estimate is confirmed by a more careful analysis of the open-boundary Green's function provided in App.~\ref{app:Steady_State_Occupation}.
       
       It is demonstrated that  $G^R_{\rm obc}[j,p;\omega] \approx G^{R}_{\infty}[j-p;\omega]$ and $G^R_{\rm obc}[j,p;0] \sim e^{\mp|j-p|/(w/\Gamma)}$ are excellent approximations in the strong non-reciprocity $\kappa \approx w$ and weak pumping $\Gamma \ll w, \kappa$ limit.  Generically, one can also associate a (direction-dependent) decay length to $G^{R}[j,p;w]$ for arbitrary $\omega$: for $j > p$ it is roughly $w/\Gamma$ near zero-frequency and drops quickly to zero outside the band $\omega \approx w$.  However in the limit of interest where the spacing of the modes is much smaller than their width, the occupation is dominated by low frequencies and we find (see App.~\ref{app:Steady_State_Occupation} for details)
       \begin{align} \nonumber
	        \langle \hc_j^\dagger \hc_{j} \rangle_{\rm ss}
	        &\stackrel{\kappa \to w}{\approx}
	        \frac{\Gamma}{\kappa\pm \Gamma}
	        \left(
	        1
	        +
	        C_1
	        \sum_{p \leq j}
	        \frac{e^{\mp2 |j-p|/(w/\Gamma)}}{\sqrt{|n-l|}}
	        \right)
	        \\ \label{eq:Steady_State_Occupation_Full_Gamma}
	        &\approx
	        \begin{cases}
	        \frac{\Gamma}{\kappa+\Gamma}
	        \left(
	        \frac{\text{erf}(\sqrt{\frac{j}{\xi_{\rm obc}}})}{\sqrt{\xi_{\rm obc}^{-1}}} + C_2
	        \right), 
	        & \text{fermions}
	        \\
	        \frac{\Gamma}{\kappa-\Gamma}
	        \left(
	        \frac{\text{erfi}(\sqrt{\frac{j}{\xi_{\rm obc}}})}{\sqrt{\xi_{\rm obc}^{-1}}} + C_3
	        \right), 
	        & \text{bosons}.
	        \end{cases}
	    \end{align}
       In the first line, we immediately consider the limit of extreme non-reciprocity where only pump baths attached to sites on the left of a given site $j$ contribute to the occupation. In this limit, the localization length $1/A$ associated with the NHSE goes to zero yet, in contrast, we see the emergence of a new length scale $\xi_{\rm obc} = w/(2\Gamma)$ which controls the quantum steady state of the open boundary condition system. Here, $\text{erf}$ and $\text{erfi}$ are the error and imaginary error function respectively, 
        and $C_1,C_2,C_3$ are constants (see App.~\ref{app:Steady_State_Occupation}). 


        
        Equation (\ref{eq:Steady_State_Occupation_Full_Gamma}) is a central result of our work. It demonstrates that the spatial distribution of the steady-state occupation is controlled by a new macroscopic length scale $\xi_{\rm obc} = w/(2\Gamma)$, despite the absence of a long time scale. In addition, $\xi_{\rm obc}$ does not resemble any right or left eigenvector of $\Heff$. Colloquially, the open-boundary steady state has ``remembered'' the long propagation length of the periodic chain but not the associated long time scale that came with it. Further, it shows that this new emergent lengthscale  has a dramatically different interpretation depending on particle statistics. For fermions it corresponds to a \textit{decay length}, whereas for bosons it represents an \textit{amplification length}. 
        
        In Fig.~\ref{fig:Model_Steady_State_Occupation} we plot the steady-state occupation $\langle \hc_j^\dagger \hc_j \rangle_{\rm ss}$ and find behaviour consistent with the above interpretation. The density of fermions in the steady state is suppressed over a length scale $\xi_{\rm obc}$ on the left edge of the system, and saturates as we move to the right at a value coinciding with the expected constant PBC occupancy. In contrast, the number of bosons increases exponentially as one moves to the right edge of the system, with $\xi_{\rm obc}$ now serving as the associated localization length. The numerically-computed results in Fig.~\ref{fig:Model_Steady_State_Occupation} do not differ from the analytical prediction of Eq.~(\ref{eq:Steady_State_Occupation_Full_Gamma}) on the presented scale. 

        In Fig.~\ref{fig:Steady-State_Occupation_Vary_Gamma} we plot the scaled steady-state occupation $\langle \hc_j^\dagger \hc_j \rangle_{\rm ss}/\langle \hc_N^\dagger \hc_N \rangle_{\rm ss}$ over a range of $\Gamma$.  The behaviour here has a simple origin.   Vanishingly small pumping $\Gamma \approx 0$ implies a large length scale $\xi_{\rm obc} \gg N$ and we expect the density of particles in the steady-state to be essentially independent of statistics: the particles must propagate a large distance before the effects of $\Gamma$ are noticeable. This is verified by taking the $\xi_{\rm obc} \gg j$ limit of Eq.~(\ref{eq:Steady_State_Occupation_Full_Gamma}), which gives
        \begin{align} 
	    \langle \hc_j^\dagger \hc_{j} \rangle_{\rm ss}
	    & \sim
	    \frac{\Gamma}{\kappa\pm \Gamma}
	    \left(
	     1+\frac{2}{\sqrt{\pi}}\left(\sqrt{j}-1 \right)
	    \right)
        \end{align}
        where the $+$ ($-$) sign is for fermions (bosons). 
        
        As we increase the pumping, fermions will experience additional damping. Particles injected by the baths into the chain can therefore not propagate as far before decaying. The length scale $\xi_{\rm obc}$ thus becomes smaller, and the particle density will saturate more
        rapidly as we move from left to right.  In contrast, bosons are less damped as we increase $\Gamma$, but this simply means particles can propagate further, and we expect that the accumulation of particles on the right edge becomes more pronounced. Both these predictions are in agreement with Fig.~\ref{fig:Steady-State_Occupation_Vary_Gamma}, and with the analytical result obtained for $\xi_{\rm obc}$ in the limit of larger pumping  (see Eq.~(\ref{app:xi_obc_General})).


  
	   The analysis presented in this section shows conclusively that in a quantum realization of the Hatano-Nelson model, a large dissipative gap in the spectrum of $\Heff$ does not preclude the existence of a long length scale. Further, neither the spatial structure of left or right eigenvectors can be used to infer the steady-state occupation $\langle \hc^\dagger_j \hc_j \rangle_{\rm ss}$ under open boundary conditions. Do these two features holds generically? Referring back to Eq.~(\ref{eq:Real_Space_Occupation}) and the accompanying discussion, we expect this to be the case. The steady-state occupation is controlled by the real-space Green's function and unlike the spectrum and eigenvectors, the former is largely unaffected by the NHSE when changing boundary conditions. The long attenuation length in the infinite-sized model can therefore show up in the finite open-boundary case, despite having a large damping gap. To verify this prediction, we briefly analyze two additional non-Hermitian models in App. \ref{app:Two_More_Models}. Both models exhibit, a large dissipative gap, yet there stills exists a large length scale in both models which can be extracted from the Green's function of the infinite-sized model.   
	   
	   The validity of $G^R_{\rm obc}[j,p;\omega] \approx G^{R}_{\infty}[j-p;\omega]$ also makes it clear that the sensitivity to boundary conditions is not caused by the NHSE. Once this approximation has been made, the only remaining information regarding boundary conditions is the trivial one: particles can not tunnel directly between the first and last site. Given the propensity for particles to propagate to the right, this implies that a pump bath attached to the first site can contribute to to the total particle number on site $N$, but not vice-versa. Thus, the asymmetry we see in the occupation is categorically not due to a change in the spectrum or eigenvectors of $\Heff$ and therefore not due to the NHSE. Rather, the built-in non-reciprocity of the effective Hamiltonian is the sole reason we see just a drastic change under different geometries.

		\begin{figure}[t]
		\centering
		\includegraphics[width=0.475
		\textwidth]{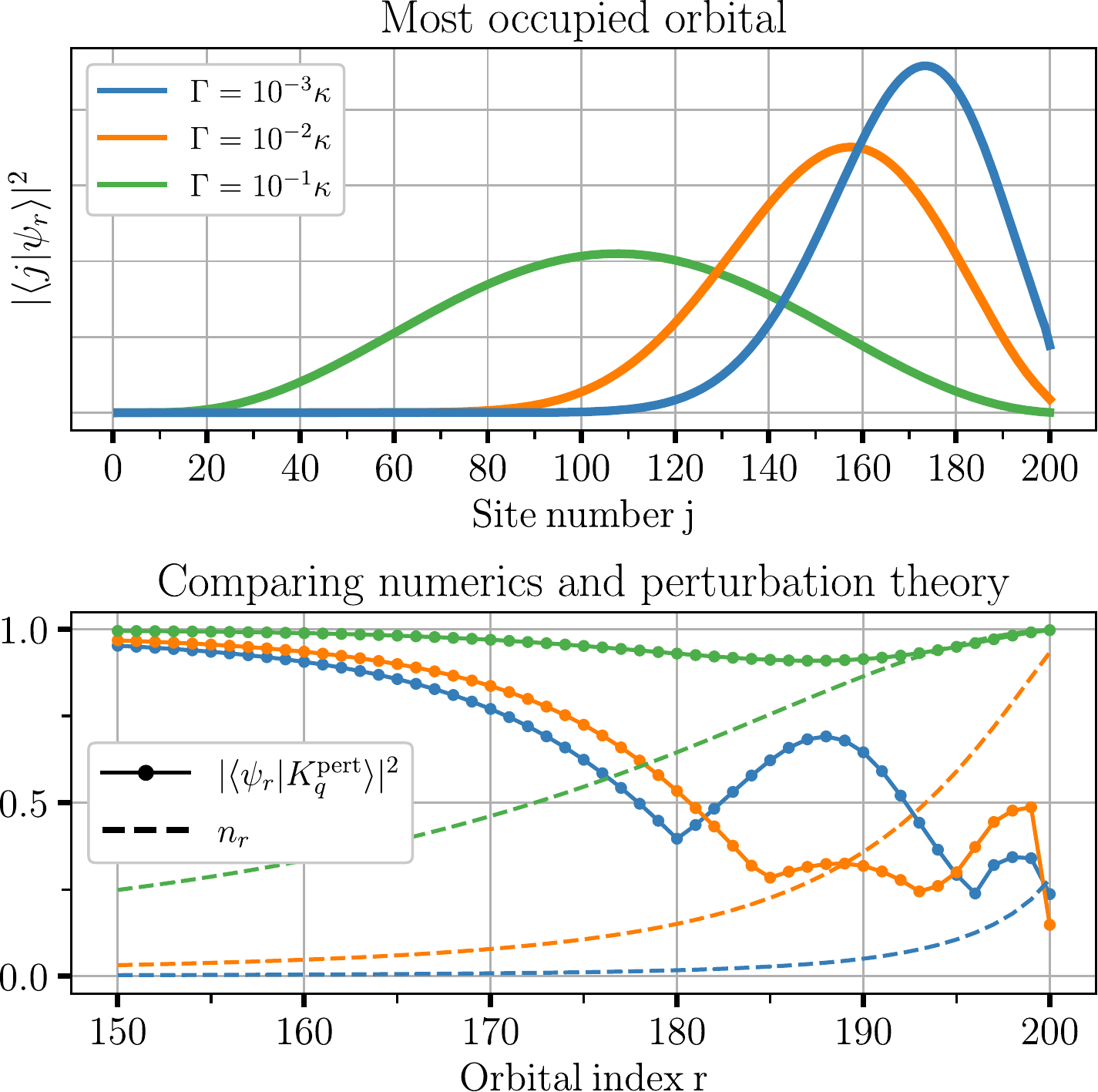}
		\caption{
		Top: real-space wavefunction squared $|\braket{j}{\psi_{r}}|^2$ of the most occupied steady state orbital for our quantum Hatano-Nelson model with open boundary conditions, $N = 200$ and 
		$w = 0.9 \kappa$. The model is realized using a coherent Hamiltonian and dissipators Eqs.~(\ref{eq:H_coh_HN})-(\ref{eq:G_n_HN}). 
		Despite the strong non-reciprocity for our chosen parameters, this dominant orbital wavefunction does not exhibit the exponential localization associated with the NHSE.  
		Bottom:  same parameters, steady state orbital occupation $n_r$ (dashed lines) and the overlap squared $|\langle \psi_r | K_{q}^{\rm pert} \rangle|^2$ between the exact orbitals $\ket{\psi_r}$ and the ones obtained from second order perturbation theory in $w$,  $\ket{K_{q}^{\rm pert}}$ (dotted lines).  Despite the large value of $w$, the perturbative expression for the orbitals still provides a reasonable approximation for most orbitals.  This implies that the majority of orbitals are spatially extended over the whole chain, and have no resemblance to the right and left eigenvectors of $\Heff$. 
            }
		\label{fig:Overlap__Squred_Perturbation}
	    \end{figure}
	 
		\subsection{Orbitals}
	    The steady-state real-space average occupation only gives us partial information about $\hrho_{\rm ss}$. In general, one would need knowledge of all possible steady-state correlations to recreate $\hrho_{\rm ss}$. The Gaussian nature of our steady state greatly simplifies this task: all higher-order correlations are completely determined by the two-point correlation matrix $\boldsymbol{F}$ (i.e.~Wick's theorem holds). 
	    Indeed, if we let $\ket{\psi_r}$ denote the eigenvectors of $\boldsymbol{F}$ with eigenvalue $n_{r}$ then the Gaussian nature of the steady state implies
	    \begin{align}
	        \hrho_{\rm ss}
	       =
	       \frac{ \exp
	        \left(
	        \sum_{r}
	        \ln(\frac{n_r}{1\mp n_r})
	        \hc_r^\dagger \hc_r
	        \right)}
	        {\mathcal{Z}}
	    \end{align}
	    where $\mathcal{Z}$ is a normalization constant, $-$ ($+$) is for fermions (bosons) and $\hc_r$ and $\hc_r^\dagger$ are the (orthogonal) annihilation and creation operators associated with $\ket{\psi_r}$.
	    We thus see that the eigenvectors $\ket{\psi_r}$ represent real-space wavefunctions of orbitals that are independently occupied in the steady state.   
	    
	    Identifying this structure leads to an obvious question:  does the spatial structure of these occupied-orbital wavefunctions change drastically in going from periodic to open boundary conditions?  Note that these wavefunctions $\ket{\psi_r}$ are eigenvectors of a Hermitian matrix, and hence form a complete orthonormal basis.  They thus cannot coincide with either the right or left eigenvectors of $\Heff$.  In what follows, we show something even stronger:  these occupied orbital wavefunctions do not exhibit any singular behaviour remotely analogous to the non-Hermitian skin effect.  
	    
	    It is tempting to argue that this is to be expected, given that $\boldsymbol{F}$ is a Hermitian matrix. One might argue that these objects, unlike their non-Hermitian counterparts, are more robust to perturbations (such as varying boundary conditions) and that singular behavior of their eigenstates is ruled out by construction. This intuition is however incorrect. Perhaps the best way to demonstrate why this reasoning is incorrect is to express $\boldsymbol{F}$ as
	    \begin{align}
	        \boldsymbol{F}
	        =
	        \int_{0}^{\infty}
	        dt
	        e^{-i \Heff t}
	        \boldsymbol{G}
	        e^{i  \Heff^\dagger t}.
	    \end{align}
	    Perturbations to $\Heff$ which are naively thought to be small therefore have an infinite amount of time to influence the steady state. This can already be seen by considering the steady-state occupation $\langle \hc_j^\dagger  \hc_j \rangle_{\rm ss}$ of fermions versus bosons as see in Fig.~(\ref{fig:Model_Steady_State_Occupation}). The only difference between their effective Hamiltonians is the small uniform rate $\Gamma$. Nevertheless, this leads to an exponentially large difference in average particle number on the last site $\langle \hc_N^\dagger \hc_N \rangle_{\rm ss} \sim e^{\mp 2 \Gamma/w N}$. It it thus not immediately obvious that changing boundary conditions does not drastically alter the orbitals $\ket{\psi_r}$ in analogy with the NHSE. 

		
		We focus on the most interesting regime of strong non-reciprocity
		where $w$ is just slightly smaller than $\kappa$.  Obtaining the desired occupied-orbital wavefunctions $\ket{\psi_r}$ analytically by diagonalizing $\boldsymbol{F}$ is unfortunately infeasible.  Instead, we use a perturbative argument that has three steps.  We first find the steady-state correlation matrix $\boldsymbol{F}_0$ for the case where the coherent hopping vanishes $w \to 0$.  We next calculate the leading order corrections for small $w$.  Finally, we argue that (somewhat surprisingly) this perturbative-in-$w$ expression remains approximately valid even for  modestly large values of hopping.  
		
		Consider then the case $w = 0$, where we have a fully reciprocal system but with only dissipative hopping processes mediate by the loss baths.  
        One finds that in this case, the steady state covariance matrix has the form
	    \begin{align}
	       \boldsymbol{F}_0
	       =
	       \sum_{K_q}
	       n_{K_q}
	       \ket{K_q}
	       \bra{K_q}.
	    \end{align}
	    Here $\ket{K_q}$ are standing-wave states with a center of mass momentum $\pi/2$
	    \begin{align}
	        \braket{j}{K_q}
	        =
	        \sqrt{\frac{2}{N+1}}
	        e^{i \frac{\pi}{2}j}
	        \sin K_q j.
	    \end{align}
	    $n_{K_q}$ is the corresponding occupation, completely determined by the eigenvalues of $\boldsymbol{G}$ and $\boldsymbol{L}$
	    \begin{align} \label{eq:n_K} 
	        n_{K_q}
	        &
	        =
	        \frac{G_{K_q}}{L_{K_q}+G_{K_q}}
	        =
	        \frac{\Gamma}{\kappa(1+\cos K_q) + \Gamma}
	    \end{align}
	    with $K_q = q\pi /(N+1)$ and $q$ an integer that runs from $1$ to $N$. The orbitals and occupation numbers strongly resemble what we would find with periodic boundaries. This should not be a surprise, as there is no non-reciprocity and all the dynamics can be understood in terms of the Hermitian matrix $\boldsymbol{L}$.
	    
	   Now we wish to consider the effect of a finite but small coherent hopping matrix element $w$. Writing the steady-state correlation matrix as $\boldsymbol{F} = \boldsymbol{F}_0+w \boldsymbol{F}_1+\mathcal{O}(w^2)$ and using Eq.~(\ref{eq:Lyaponov}) we obtain 
	    \begin{align}\label{eq:First_Order_Correction}
	        \langle
	        K_{q}
	        |
	        \boldsymbol{F}_1
	        |
	        K_{q'}
	        \rangle
	        =
	        \frac{i}{w}
	        \frac{n_{K_q}-n_{K_{q}'}}
	        {G_{K_q}+L_{K_q}+G_{K_{q'}}+L_{K_{q'}}}
	 	 \langle
	        K_{q}
	        |
	        \boldsymbol{H}
	        |
	        K'_{q}.
	        \rangle
	    \end{align}
	   The overlap $\langle K_{q} | \boldsymbol{H} | K_{q'} \rangle$ is zero if $K_q$ and $K_{q'}$ have the same parity and
	    \begin{align} \label{eq:Overlap}
	        \langle K_{q} | 
	        \boldsymbol{H}
	        | K'_{q} \rangle
	        &=
	        \frac{w i}{N+1}
	        \frac{\sin K_q \sin K_q' }{\sin(\frac{K_q'-K_q}{2})\sin(\frac{K_q'+K_q}{2})}
	        \\ \label{eq:Overlap_Two_Extremes}
	        &\approx
	        \begin{cases}
	       \frac{w i}{N+1}, & K_q\approx K_{q'} \approx 0 \: \text{or} \: \pi
	       \\
	       \frac{w 2i}{\pi}, & K_q \approx K_{q'} \approx \frac{\pi}{2}
 	        \end{cases}
	    \end{align}
	    otherwise. In addition to the expected overlap $\langle K_{q} | \boldsymbol{H} | K_{q'} \rangle$ there are two factors that contribute to the first order correction. The numerator of Eq.~(\ref{eq:First_Order_Correction}) implies that states which have nearly identical occupations $n_{K_q} \approx n_{K_{q'}} \implies K_q \approx K_{q'}$ are not strongly mixed. States which are weakly damped $G_{K_q}+L_{K_q} \approx \Gamma \implies K_q \approx \pi$ are more sensitive to the perturbation than states which are heavily damped, which occurs when $K_q \approx 0$. We stress that Eq.~(\ref{eq:First_Order_Correction}) implies that the smallest dimensionless parameter in our problem is $w/\Gamma$ (i.e. when $K_q \approx K_{q'} \approx -\pi$).
	    
	    The higher-order perturbative corrections to $\boldsymbol{F}$ have a similar structure to Eq.~(\ref{eq:First_Order_Correction}) in that they also depend on the ratio of population differences to decay rates of the unperturbed orbitals (see App.~\ref{app:Pert_Theory_Orbitals}). Taking all three terms into account, we thus expect orbitals near $K_q \approx 0$ to be nearly unchanged by the perturbation $\boldsymbol{H}$ even when $w \approx \kappa$, both because their damping rate is large, and the overlap $\langle K_{q} | \boldsymbol{H} | K_{q'} \rangle$ with other modes is small. Conversely, we expect orbitals near $K_q \approx \pi$ to mix strongly since they are weakly damped. 
	    
	    The above expectation is borne out by comparing the numerically computed orbitals $\ket{\psi_r}$ and the second-order correction $\ket{K_q^{\rm pert}}$ to $\boldsymbol{F}$, as shown in Fig.~\ref{fig:Overlap__Squred_Perturbation}. Despite being in a regime with strong non-reciprocity and hence large $w$ ($w = 0.9 \kappa$), perturbation theory still describes the 150 least-occupied orbitals (which we do not plot) extremely well: they are nearly standing-wave states. The most-occupied orbitals are the ones which perturbation theory describes less accurately.  Nonetheless, these orbitals are still relatively delocalized, and have no resemblance to the right or left eigenvectors of $\Heff$. Finally as we increase $\Gamma$, and thus the decay rate of all modes, our perturbation theory is more accurate, as predicted by Eq.~(\ref{eq:First_Order_Correction}).
	
	    These results verify the central claim of this section:  the occupied orbitals which characterize our quantum steady state do not exhibit any analogue of the NHSE.  These occupied orbitals have a form similar to the reciprocal case $w=0$, and do not exhibit any exponential localization.

    \section{Conclusion}
    
    We have presented a systematic method for constructing open quantum systems whose unconditional physics reflects that of a desired non-Hermitian tight-binding lattice Hamiltonian.  We identify generic features of the steady states of such quantum lattice models, focusing on the case where the target Hamiltonian is non-reciprocal.  A crucial conclusion is that the steady state cannot in general be understood solely using the effective Hamiltonian $\hH_{\rm eff}$.
    First, fluctuations play a crucial role in determining $\hrho_{\rm ss}$, and their form is not uniquely determined by $\hH_{\rm eff}$ (though is constrained by it). Further, taking the Hatano-Nelson model as an example, we have demonstrated that even when fluctuations are spatially featureless (i.e.~uniform incoherent pumping), the spectrum, left and right eigenvectors of $\hH_{\rm eff}$ cannot be simply used to infer even the most basic features of the steady state. 
    For weak pumping, we find that our system under open boundary conditions exhibits a new long length scale $\xi_{\rm obc}$.  This scale has no relation to the localization length of the left and right eigenvectors, nor to the existence of an extremely small relaxation rate (dissipative gap). Particle statistics also play a surprising role in the form of the steady state. Finally, we have shown that the orbital states (the eigenstates of $\hrho_{\rm ss}$) do not exhibit any analogue of the non-Hermitian skin effect.  Unlike the left and right eigenvectors of $\hH_{\rm eff}$, they do not become exponentially localized under a change of boundary conditions; in fact, a majority of them are very nearly standing-wave states. 
    
    Our work naturally leads to several interesting questions regarding the interplay of dynamics, noise, and the steady state. For instance, one could ask to what extent non-trivial correlated fluctuations (i.e.~when the incoherent pumping has a non-trivial spatial structure) can lead to an interesting steady state. To that end, we briefly analyze such a model in App.~\ref{app:G_Not_Identity}, where we find a steady-state occupation which might naively be interpreted as a consequence of the NHSE, despite the lack of non-reciprocity. Taking the opposite approach presented in this paper, by starting with a target steady state and attempting to classify the allowable $\boldsymbol{H}_{\rm eff}$ and $\boldsymbol{G}$, would also be extremely interesting.
    \section*{Acknowledgements}
    
    This work was supported by the Army Research Office under grant W911-NF-19-1-0380.  AC acknowledges support of the Simons Foundation through a Simons Investigator award.
    
    \appendix

\section{Equations of motion - correlation functions}\label{app:Normal_Order_Corr}
In this appendix, we derive the equations of motion for the normal ordered correlation function $\langle \hc_n^\dagger \hc_m \rangle $. The jump operators are $\hat{L}_{\mu} = \sum_n l_{\mu m} \hat{c}_m$ and $\hat{G}_{\nu} = \sum_{n} g^*_{\nu n} \hc_n^\dagger$, from which we obtain
\begin{align} \nonumber
    i \partial_t \hrho
    &
    =
    \sum_{a,b}
    H_{ab}[ \hc_a^\dagger \hc_b, \hrho ]
    +
    i\sum_{a,b}
    L_{ab} 
    \left(
    \hc_b \hrho \hc_a^\dagger
    -\frac{1}{2}
    \{\hc_a^\dagger \hc_b , \hrho\}
    \right)
    \\
    &
    +
    i \sum_{a,b}
    G_{ab} 
    \left(
    \hc_a^\dagger \hrho \hc_b
    -\frac{1}{2}
    \{\hc_b \hc_a^\dagger , \hrho\}
    \right)
\end{align}
where $\{\cdot, \cdot \}$ is the anti-commutator and, as in the main text, $L_{ab} = (\boldsymbol{l}^\dagger \boldsymbol{l})_{ab}$ and $G_{ab} = (\boldsymbol{g}^\dagger \boldsymbol{g})_{ab}$. Thus, the equations of motion for the normal ordered correlation function are
\begin{align}\nonumber
   i \partial_{t} \langle \hc_n^\dagger \hc_m  \rangle &
   =
   \sum_{a,b}
   H_{a,b}
   \langle
   [\hc_n^\dagger \hc_m,\hc_a^\dagger \hc_b]
   \rangle
   \\ \nonumber
   &
   +
   \frac{i}{2}\sum_{a,b}
   L_{ab}
   \langle
   \hc_a^\dagger
   [\hc_n^\dagger \hc_m, \hc_b ]
   +
   [\hc_a^\dagger, \hc_n^\dagger \hc_m]
   \hc_b
   \rangle
   \\
   &
   +
   \frac{i}{2}\sum_{a,b}
   G_{ab}
   \langle
   \hc_b
   [\hc_n^\dagger \hc_m, \hc_a^\dagger ]
   +
   [\hc_b, \hc_n^\dagger \hc_m]
   \hc_a^\dagger
   \rangle   
\end{align}
 Using fermionic anti-commutation and bosonic commutation relations, we get
 \begin{align} \nonumber
       i \partial_{t} \langle \hc_n^\dagger \hc_m  \rangle &
   =
   \sum_{a}
   \left(
   H_{ma}
   \langle
   \hc_n^\dagger \hc_a
    \rangle
    -
    H_{an}
   \langle
   \hc_a^\dagger \hc_m
    \rangle
    \right)
   \\ \nonumber
   &
   -
   \frac{i}{2}
  \sum_{a}
   \left(
   L_{ma}
   \langle
   \hc_n^\dagger \hc_a
    \rangle
    +
    L_{an}
   \langle
   \hc_a^\dagger \hc_m
    \rangle
    \right)
   \\
   &
    \mp
   \frac{i}{2}
  \sum_{a}
   \left(
   G_{ma}
   \langle
   \hc_a \hc_n^\dagger
    \rangle
    +
    G_{an}
   \langle
   \hc_m \hc_a^\dagger 
    \rangle
    \right)
 \end{align}
 where $\mp$ is for fermions and bosons respectively. Using the anti-commutation and commutation relation once more, along with $\boldsymbol{L}^\dagger = \boldsymbol{L}$ and $\boldsymbol{G}^{\dagger} = \boldsymbol{G}$, we recover Eq.~(\ref{eq:EOM_Correlation}) in the main text. The anti-normal-ordered correlation function has the same dynamical matrix $\boldsymbol{H}_{\rm eff}$, but with a different inhomogeneous term:
    \begin{align} \nonumber
        i\partial_t \langle \hc_m \hc_n^\dagger \rangle
        &=
        \sum_{a}
        \left(
        (H_{\rm eff})_{ma} \langle \hc_a \hc_n^\dagger  \rangle
        -
        (H_{\rm eff}^\dagger)_{an}\langle \hc_m  \hc_a^\dagger\rangle
        \right)
        \\ 
        &+ i L_{mn},
    \end{align}
    which follows from the preservation of equal time anti-commutation or commutation relations.
    
\section{Hatano-Nelson Hamiltonian using Method 2}\label{app:HN_Method_2}
Here we construct a quantum effective non-Hermitian Hamiltonian and noise matrix whose dynamics are equivalent to a Lindblad master equation for the Hatano-Nelson model using Method 2 in the main text. Writing the target Hamiltonian Eq.~(\ref{eq:HN_First_Quantized}) in momentum space, we have
\begin{align}
    \hat{H}_{\rm targ}^{\rm HN}
    =
    \sum_{k}
    \left(
    w \cos k- i \kappa \sin k 
    \right)
    \hc_k^\dagger \hc_k
\end{align}
The largest positive eigenvalue of the anti-Hermitian part is clearly $\kappa$. Thus, the effective Hamiltonian reads
\begin{align}
   \hat{H}_{\rm eff}
    =
    \sum_{k}
    \left(
    w \cos k- i \kappa(1+\sin k)\mp i \Gamma 
    \right)
    \hc_k^\dagger \hc_k
\end{align}
where, as always, $\mp$ is for fermions and bosons respectively so that $\boldsymbol{H}_{\rm eff} = \boldsymbol{H}^{\rm HN}_{\rm targ}-i(\kappa\pm \Gamma) \boldsymbol{1}$. To obtain a set of real-space dissipators, it is convenient to write the master equation in momentum space as in Eq.~(\ref{eq:EOM_Momentum}) then Fourier transform back to position space. We obtain
	\begin{align} \nonumber
	i \partial_t \hrho
    		&=
			\frac{w}{2}
			\sum_{j}
			[
			\hc_{j+1}^\dagger\hc_j+h.c.  
			,
			\hrho
			]
			\\  \label{eq:EOM_Real_Space_App}
			&
			+
			i 
			\sum_{j}
			\mathcal{D}[\sqrt{\kappa}(\hc_j-i \hc_{j+1})] \hrho 
			+
			i
			\sum_{j}
			\mathcal{D}[\sqrt{2 \Gamma}\hc_j^\dagger] \hrho
	\end{align} 
	i.e. the coherent Hamiltonian and dissipators in Eqs.~(\ref{eq:H_coh_HN})-(\ref{eq:G_n_HN}). Note that there is a small but important detail when dealing with different boundary conditions. For periodic boundary conditions $\hc_j = \hc_{j+N}$, Eq.~(\ref{eq:EOM_Momentum}) and  Eq.~(\ref{eq:EOM_Real_Space_App}) are equivalent whereas for open boundary conditions $\hc_{0} = \hc_{N+1} = 0$ they are not. For this choice of geometry, there are independent loss baths attached to site 1 and $N$ in addition to the one on each bond that lead to non-reciprocal hopping. The lack of translational invariance then precludes the possibility of writing the master equation where the jump operators are independent momentum creation and annihilation operators.  
	
    \section{Steady state correlation function - PBC}\label{app:N_ss}
    Using the definition of the periodic system lengthscale $\kappa \cosh \xi_{\rm pbc}^{-1} = \kappa+ \Gamma$, we have
    \begin{align} \nonumber
        &\langle \hc^\dagger_j \hc_p \rangle_{\rm ss}
        =
        \frac{\Gamma}{\kappa N}
        \sum_{k}
        \frac{e^{-i(j-p)k}}{\cosh \xi^{-1}_{\rm pbc}+ \sin k}
        \\ \label{eq:N_ss}
        &
        =
        \frac{\Gamma}{\kappa N \sinh \xi_{\rm pbc}^{-1}}
        \sum_k
        \left(
        \frac{e^{-i(j-p)k}}{1+i e^{-i k} e^{-\xi_{\rm pbc}^{-1}}}
        -
        \frac{e^{-i(j-p)k}}{1+i e^{-i k} e^{\xi_{\rm pbc}^{-1}}}
        \right)
    \end{align}
    Since the $k$ are integer multiples of $2\pi/N$, we can perform a simple geometric sum to obtain
    \begin{align}\label{eq:Geometric_Sum}
        \sum_{l=0}^{N-1}
        \frac{i^{-l}e^{-i k l} e^{\pm \xi_{\rm pbc}^{-1} l}}
        {1-i^{-N}e^{\pm \xi_{\rm pbc}^{-1} N}}
        =
        \frac{1}{1+i e^{-i k} e^{\pm\xi_{\rm pbc}^{-1}}}
    \end{align}
    Inserting Eq.~(\ref{eq:Geometric_Sum}) into Eq.~(\ref{eq:N_ss}), only the $l=p-j$ term survives, again due to the quantization of $k$ (where $p-j$) is understood modulo $N$. After some simplification we are left with
    \begin{align}
        \langle \hc^\dagger_j \hc_p \rangle_{\rm ss}
        =
        \frac{\Gamma i^{-(p-j)} }{\kappa \sinh \xi_{\rm pbc}^{-1}}
        \left(
        \frac{e^{-(p-j)/\xi_{\rm pbc}}}{1-i^{-N} e^{-N/\xi_{\rm pbc}}}
        -
        \frac{e^{(p-j)/\xi_{\rm pbc}}}{1-i^{-N} e^{N/\xi_{\rm pbc}}}
        \right)
    \end{align}
    which recovers the scaling of Eq.~(\ref{eq:Avg_Particle_PBC_2}) in the main text. 
    
    \section{Steady-state occupation under open boundary conditions}\label{app:Steady_State_Occupation}
    In this appendix, we find approximate expressions for the steady-state occupation $\langle \hc_j^\dagger \hc_j \rangle_{\rm ss}$ under open boundary conditions. It is worth pointing out immediately that since the eigenvectors and eigenvalues of the Hatano-Nelson model are known \cite{Hatano_Nelson}, we can use the formal expression Eq.~(\ref{eq:F_Right_Left}) and solve for $\langle \hc_j^\dagger \hc_j \rangle_{\rm ss}$ without approximations. Assuming $w > \kappa$ throughout, we have
    \begin{align}\label{eq:Exact_Occupation}
    \langle \hc_j^\dagger \hc_j \rangle_{\rm ss}
        =
        \sum_{K_q, K_{q'}, p}
        \frac{e^{2A(j-p)}}{\mathcal{N}}
        \frac{\sin K_q j \sin K_q p \sin K_{q'}p \sin K_{q'}j }{J (\cos(K_q)-\cos(K_{q'})) - i2(\kappa \pm \Gamma)}
    \end{align}
    where $\mathcal{N}^{-1} = -i8 \Gamma/(N+1)^2 $, $2A = \ln((w+\kappa)/(w-\kappa))$, and $J = \sqrt{w^2-\kappa^2}$. The standing-wave momenta are quantized by $K_q = \pi q/(N+1)$ where $q$ is an integer that runs from 1 to $N$. Although exact, Eq.~(\ref{eq:Exact_Occupation}) is practically useless, and does not immediately tell us qualitative features of the steady-state occupation. For instance, it is not evident that the small difference in uniform dissipation for fermions and bosons $\kappa \pm \Gamma$ can lead to a drastic change in $\langle \hc_j^\dagger \hc_j \rangle_{\rm ss}$ as seen in Fig.~\ref{fig:Model_Steady_State_Occupation}.
    
    We will instead work with the other formal solution to the steady-state occupation, given by Eq.~(\ref{eq:Real_Space_Occupation}). We must therefore first find the retarded frequency-space Green's function of the open chain. This has been done in previous work \cite{Alexander_PRX, Alexander_Nat_Comm}, but we present here a different approach that will allow us to simultaneously find the response of the periodic chain and compare in what manner the two differ. To that end, we first find the time-domain Green's function $G^{R}_\infty(j,p;t)$ for an infinite lattice, whose equations of motion are
    \begin{align} \nonumber
        &i \partial_t 
        G^R_{\infty}(j,p;t)
        -
        \frac{(w+\kappa)}{2}G^R_{\infty}(j-1, p; t)
        \\ \label{eq:GF_EOM}
        &
        -
        \frac{(w-\kappa)}{2}G^R_{\infty}(j+1,p;t)
        +is_{\pm} G^R_{\infty}(j,l;t)
        =
        \delta(t) \delta_{j,p}.
    \end{align}
    along with the initial condition $G^R_\infty(j,p,0) = -i \delta_{j, p}$
    with $\delta(t)$ the Dirac delta function and $\delta_{j,p}$ the Kronecker delta function. For notational convenience, we have set $s_{\pm} = \kappa \pm \Gamma$. These can readily be solved by using the plane wave solutions, and with an infinite-sized system there is no quantization condition on the momentum. Thus
    \begin{align} \nonumber
        G^R_\infty(j,p;t)
        &
        =
        -i
        \Theta(t)
        e^{-s_{\pm}t}
        \int_{-\pi}^{\pi}
        \frac{d k}{2\pi}
        e^{ik(j-p)}
        e^{-i (w \cos k-i \kappa \sin k) t}
        \\ \label{eq:G_infty}
        &
        =
        -i
        \Theta(t)
        e^{-s_{\pm}t}
        e^{A(j-l)}
        \int_{-\pi}^{\pi}
        \frac{d k}{2\pi}
        e^{ik(j-p)}
        e^{-i J \cos k t}
    \end{align}
    where in the last line we have made an imaginary gauge transformation $k \to k -i A$ using the definition of Eq.~(\ref{eq:Def_A}). Although seemingly ad-hoc, this transformation can be justified using complex analysis. Viewed as a complex variable, the integrand is a holomorphic function of $k$, and one then constructs a rectangle in the complex plane over which the integral is zero. Two sides of the rectangle cancel, after which one equates the two functions above.
    
    The frequency-space response is obtained by taking the Fourier transform of Eq.~(\ref{eq:G_infty}), after which we have 
    \begin{align}\nonumber
        G^{R}_{\infty}[j,p;\omega]
        &
        =
        e^{A(j-p)}
        \int_{-\pi}^{\pi}
        \frac{dk}{2\pi}
        \frac{e^{i k(j-p)}}{w+i s_{\pm}-J \cos k}
        \\
        &
        =
        -i
        \frac{e^{A(j-p)}e^{-iQ[\omega]|j-p|}}
        {J \sin Q[\omega]}
    \end{align}
    where we have used the residue theorem to compute the integral. The complex wavevector $Q[\omega]$ is defined by
    \begin{align}\label{eq:Dispersion}
    w+ i s_{\pm} = J \cos Q[\omega]
    \end{align}
    and we always chose the imaginary part of $Q[\omega]$  to be negative such that $e^{-i Q[\omega]}$ lies in the unit circle. One can also readily verify that this satisfies the Fourier-transformed version of Eq.~(\ref{eq:GF_EOM}).
    
    To obtain the retarted response of a chain with periodic boundary conditions, we note that the equations of motion Eq.~(\ref{eq:GF_EOM}) 
    remain unchaged, except the boundary conditions are now $G^{R}_{\rm pbc}[j+N,p; \omega] = G^{R}_{\rm pbc}[j,p+N; \omega] = G^{R}_{\rm pbc}[j,p; \omega] $. By linearity of the equations, the solution is then
    \begin{align} \nonumber
        &
        G^R_{\rm pbc}[j,p;\omega]
        =
        \sum_{r = -\infty}^\infty
        G^{R}_\infty[j-p+Nr;\omega]
        \\ \label{eq:GR_PBC}
        &
        =
        \frac{-i e^{A(j-p)}}
        {J \sin Q[\omega]}
        \left(
       \frac{ e^{-iQ[\omega](j-p)}}{1-e^{A N-i Q[\omega] N}}
       -
       \frac{ e^{iQ[\omega](j-p)}}{1-e^{A N+i Q[\omega] N}}
        \right)
    \end{align}
    where $j-p$ is understood to be modulo $N$. Each term $r \neq 0$ in Eq.~(\ref{eq:GR_PBC}) should be interpreted as an additional round trip by the particle. That is, in going from site $p$ to $j$, the particle can propagate clockwise or counter-clockwise any number of times. The total Green's function is then the sum of each of these processes. 
    
    Next we compute the Green's function of a finite-sized open chain. The particle will propagate in the bulk as it would in the infinite-sized system, without knowledge of the boundaries. Instead of round trips as in the periodic system, however, when it reaches the edge the particle can now ``bounce'' off on an open boundary, acquiring a phase shift of $\pi$ in the process. Summing over all possible bounces, we have
    \begin{widetext}
	\begin{align} \nonumber
        G^{R}_{\rm obc}[j,p;\omega]
        &=
        \frac{-i e^{A(j-p)}}
        {J \sin Q[\omega]}
        \sum_{r = -\infty}^{\infty}
        \left(
        e^{-iQ[\omega](|2(N+1)r+|n-p||)}
        -
        e^{-iQ[\omega](|2(N+1)r+|n+p|)|}
        \right)
        \\ \label{eq:GR_Bounce}
        &=
        e^{A(j-p)}
        \frac{2 \sin Q[\omega] \min(j,p) \sin Q[\omega](N+1-\max(j,p))}{J \sin Q[\omega] \sin Q[\omega](N+1) }
	\end{align}
	\end{widetext}
	in agreement with previously obtained results \cite{Alexander_PRX}. We stress once again that each term in the sum above should be interpreted as scattering off either boundary.
	
	Our remaining task is to compute the integral over all frequency
    \begin{align}
	 \langle \hc_j^\dagger \hc_j\rangle_{\rm ss}
	 &
	 =
	 2 \Gamma 
	 \sum_{p=1}^N
	 \int_{-\infty}^{\infty}
	 \frac{d \omega}{2\pi}
	 |G^{R}_{\rm obc}[j,p;\omega]|^2.
	 \end{align}
	 We stress again that this can in principle be computed exactly, see Eq.~(\ref{eq:Exact_Occupation}). The first step in finding a more enlightening approximate expression is to note that the dominating contribution to the integral occurs when the imaginary part of the complex wavevector $Q[\omega]$ is smallest. Since the real and imaginary part of $Q[\omega]$
	 \begin{align}\label{eq:Real_Imag}
	     Q[\omega]
	     =
	     k[\omega]
	     +
	     i R[\omega]
	 \end{align}
	 are not independent, it will be convenient to first make a change of variables $\omega \to k$ in the integral. Comparing the dispersion Eq.~(\ref{eq:Dispersion}) and the definition of Eq.~(\ref{eq:Real_Imag}), while also requiring that $R[k]$ is negative, we obtain
	 \begin{align}
	 \frac{d \omega}{|\sin Q[\omega]|^2}
	 =
	 \frac{J^2 dk}{\sqrt{s_{\pm}^2 + (J\sin k)^2}}
	 \\
	 e^{R[k]}
	 =
	 \frac{\sqrt{s_\pm^2+(J \sin k)^2}-s_{\pm}}{J \sin k}
	 \end{align}
	 where the integration variable $k$ goes from 0 to $\pi$. We can interpret $-R[k]$ as a sort of momentum-dependent inverse decay length (even though we have open boundary conditions and momentum is not a good quantum number). That is, $-R[k]$ corresponds to how far can a particle with momentum $k$ can propagate in the lattice before decaying. This damping is minimal when the magnitude of the group velocity $|J \sin k|$ is maximal, i.e. at $k = \pi/2$. Conversely, when the group velocity vanishes at $k =0, \pi$, there is no propagation $-R[k] \to \infty$.
	 
	 The formal expression for the steady-state occupation then reads
	 \begin{widetext}
	 \begin{align} 
	 \langle \hc_j^\dagger \hc_j\rangle_{\rm ss}
	 &
	 =
	 8 \Gamma 
	 \sum_{p=1}^N
	 e^{2A(j-l)}
	 \int_{0}^{\pi}
	 \frac{d k}{2\pi}
	 \frac
	 {
	 |\sin Q[k] \min(j,p)|^2
	 |\sin Q[k] (N+1-\max(j,p))|^2
	 }
	 {
	 \sqrt{s_{\pm}^2+(J \sin k)^2}
	 |\sin Q[k](N+1)|^2
	 }.
	 \end{align}
	 \end{widetext}
	where $Q[k]= k + i R[k]$. We now make our first approximation and assume that the time it takes for a particle to traverse the chain, i.e. the length $N$ divided by the group velocity $J$, is much larger than its lifetime $1/s_{\pm}$. This is equivalent to the requirement in the main text that the spacing of the modes are much smaller than their widths $J/N \ll s_{\pm}$. Expanding $Q[k]$ to lowest order in $s_\pm/J$, it follows that this condition implies $|e^{-i Q[k] N}| \ll 1$ for all $k$. Consequently, we approximate the response by keeping only the leading order term in  $e^{-i Q[k]}$. From Eq.~(\ref{eq:GR_Bounce}), this is exactly equivalent to taking the no-bounce limit $G^{R}_{\rm obc}[j,p;\omega] \approx G^{R}_{ \infty}[j,p;\omega]$. We thus expect this approximation to worsen when $j$ or $p$ are near a boundary. Note that the condition $J/N \ll s_{\pm}$ is easily satisfied in the limit of strong non-reciprocity $\kappa \approx w$ since  $J = \sqrt{w^2-\kappa^2}$. 
	 
	 Within the no-bounce approximation, we obtain
	 \begin{align}
	     \langle \hc_j^\dagger \hc_j\rangle_{\rm ss}
	     \approx
	     2 \Gamma
	     \sum_{p=1}^N
	     e^{2A(j-p)}
	     \int_{0}^{\pi}
	     \frac{dk}{2\pi}
	     \frac{e^{2R[k]|j-p|}}{\sqrt{s_{\pm}^2+(J \sin k)^2}}
	 \end{align}
	 In the limit of strong non-reciprocity $J \ll s_\pm$, the $p = j$ term can be approximated as $\Gamma/s_{\pm}$. For $p \neq j$ we use Laplace's method to compute the integral. We approximate $R[k]$ by expanding to second order near its maximum at $k = \pi/2$ and compute the Gaussian integral by extending the bounds of integration to infinity. We are left with
	 \begin{align}
	 \langle \hc_j^\dagger \hc_j\rangle_{\rm ss}
	 \approx
	 \frac{\Gamma}{s_{\pm}}
	 \left(
	 1
	 +
	 \sqrt{\frac{s_{\pm}}{\pi \sqrt{s_{\pm}^2+J^2}}}
	 \sum_{p\neq j}
	 \frac{e^{2A(j-p)-2A'|j-p|}}{\sqrt{|j-p|}}
	 \right)
	 \end{align}
	 where
	\begin{align} \nonumber
	A' 
	&\equiv 
	-R\Big[\frac{\pi}{2}\Big]
	=
	\ln
	\left(
	\frac{\sqrt{s_{\pm}^2+J^2}+s_\pm}{J}
	\right)
	\\
	&=
	A+\ln\left(1\pm \frac{\Gamma}{w}\right)
	+
	\mathcal{O}
	\left(
	\frac{(w-\kappa) \Gamma^2}{w^3}
	\right)
	\end{align}
	Comparing with Eq.~(\ref{eq:Steady_State_Occupation_Full_Gamma}), we have
	\begin{align}
	    C_1
	    =
	    \sqrt{\frac{s_{\pm}}{\pi \sqrt{s_{\pm}^2+J^2}}}
	\end{align}
	In the limit of perfect non-reciprocity $\kappa \to w$ we have $A \to \infty$ and $J \to 0$. Only pump baths to the left of a given site contribute to its population and thus
	\begin{align}
	  \langle \hc_j^\dagger \hc_j\rangle_{\rm ss}
	 &\approx
	 \frac{\Gamma}{s_{\pm}}
	 \left(
	 1+
	 \frac{1}{\sqrt{\pi}}
	 \sum_{p<j}
	 \frac{e^{\mp\frac{|j-p|}{\xi_{\rm obc}}}}{\sqrt{|j-p|}}
	 \right) 
	\end{align}
	where
	\begin{align}\label{app:xi_obc_General}
	    \xi_{\rm obc}
	    =
	    \Bigg|
	    \left(
	    2
	    \ln
	    (
	    1\pm \frac{\Gamma}{w}
	    )
	    \right)^{-1}
	    \Bigg|
	\end{align}
	 Approximating the sum as an integral, we obtain
	\begin{align}
	 \langle \hc_j^\dagger \hc_j\rangle_{\rm ss}
	 &
	 \approx
	 \begin{cases}
	 \frac{\Gamma}{\kappa+\Gamma}
	 \left(
	 1
	 +
	 \frac{
	 \text{erf}
	 (\sqrt{\frac{j}{\xi_{\rm obc}}})
	 -
	 \text{erf}
	 (\frac{1}{\sqrt{\xi_{\rm obc}}})
	 }
	 {
	 \sqrt{\xi^{-1}_{\rm obc}}
	 }
	 \right)
	 ,
	 &
	 \text{fermions}
	 \\
	 	 \frac{\Gamma}{\kappa-\Gamma}
	 \left(
	 1
	 +
	 \frac{
	 \text{erfi}
	 (\sqrt{\frac{j}{\xi_{\rm obc}}})
	 -
	 \text{erfi}
	 (\frac{1}{\sqrt{\xi_{\rm obc}}})
	 }
	 {
	 \sqrt{\xi^{-1}_{\rm obc}}
	 }
	 \right)
	 ,
	 &
	 \text{bosons}
	 \end{cases}
	 \\
	 &
	 \stackrel{j \ll \xi_{\rm obc}}{=}
	 \frac{\Gamma}{\kappa \pm \Gamma}
	 \left(
	 1
	 +
	 \frac{2}{\sqrt{\pi}}
	 \left(
	 \sqrt{j}
	 -
	 1
	 \right)
	 \right).
	\end{align}
	where we recover the results in the main text by setting
	\begin{align}
	    C_2
	    =
	    1-\frac{\text{erf}(\frac{1}{\sqrt{\xi_{\rm obc}}})}{\xi_{\rm obc}^{-1}}
	    \\
	    C_3
	    =
	    1-\frac{\text{erfi}(\frac{1}{\sqrt{\xi_{\rm obc}}})}{\xi_{\rm obc}^{-1}}
	\end{align}

	We emphasize that the exact solution of Eq.~(\ref{eq:Exact_Occupation}) could not have easily predicted these analytic results. Further we have shown that bulk dynamics alone, by making the no-bounce approximation $G^{R}_{\rm obc}[j,l;\omega] \approx G^{R}_{\infty}[j,l;\omega]$, was sufficient to correctly capture the steady-state occupation.
	
    \section{Perturbation theory for orbitals}\label{app:Pert_Theory_Orbitals}
    We now use perturbation theory to find the second-order correction in $w$ to the standing-wave orbitals. This is unlike standard perturbation theory in that the steady-state correlation matrix $\boldsymbol{F}$ depends on $w$ to all orders
    \begin{align}\label{eq:F_Sum}
        \boldsymbol{F}
        =
        \sum_{r = 0}^\infty
        w^r \boldsymbol{F}_{r}.
    \end{align}
    Finding the orbitals to the correct order is then a two step process: we must first find $\boldsymbol{F}_r$ to the requisite order and only then solve for the corrected eigenstates. As discussed in the main text, the zeroeth order term is
    \begin{align}\label{eq:F_0_App}
        \boldsymbol{F}_0
        =
        \sum_{K_q}
        n_{K_q}
        \ket{K_{q}}\bra{K_{q}}.
    \end{align}
    To find $\boldsymbol{F}_r$ to any other order $r \geq 1$, we insert $\boldsymbol{F}$ into Eq.~(\ref{eq:Lyaponov}) and equate terms at each order in $w$ on both sides. We are left with a recursive relation between $\boldsymbol{F}_r$ and $\boldsymbol{F}_{r+1}$ whose solution is
    \begin{align}
        \boldsymbol{F}_{r}
        =
        \frac{i}{w}
        \sum_{K_q, K_{q'}}
        \left(
        \frac{
        \bra{K_q}
        \left(
        \boldsymbol{F}_{r-1} \boldsymbol{H}
        -
        \boldsymbol{H}\boldsymbol{F}_{r-1}
        \right)
        \ket{K_{q'}}
        }
        {L_{K_q}+G_{K_q}+L_{K_{q'}}+L_{K_{q'}}}
        \right)
        \ket{K_{q}}\bra{K_{q'}}
    \end{align}
    For instance, using Eq.~(\ref{eq:F_0_App}) we recover Eq.~(\ref{eq:First_Order_Correction}) in the main text. 
    
    Once we have computed these corrections to the steady-state correlation matrix to the desired order, we can then compute the orbitals to that same order. This proceedds more like standard perturbation theory. For instance, the unormalized corrected orbital is given by
    \begin{align} \nonumber
        \ket{K_q^{\rm pert}}
        &
        =
        \ket{K_{q}}
        +
        w
        \sum_{K_{q'} \neq K_q}
        \frac
        {
        \bra{K_{q'}}
        \boldsymbol{F}_1
        \ket{K_q}
        }
        {n_{K_q}-n_{K_{q'}}}
        \ket{K_{q'}}
        \\ \nonumber
        &
        +
        w^2
        \sum_{K_{q'} \neq K_q}
        \frac
        {
        \bra{K_{q'}}
        \boldsymbol{F}_2
        \ket{K_q}
        }
        {n_{K_q}-n_{K_{q'}}}
        \ket{K_{q'}}
        \\
        &
        +
        w^2
       \sum_{
       \mathclap{\substack{K_{q'} \neq K_q\\
                            K_{q''} \neq K_q}}}
        \frac
        {
        \bra{K_{q'}}
        \boldsymbol{F}_1
        \ket{K_{q''}}
        \bra{K_{q''}}
        \boldsymbol{F}_1
        \ket{K_{q'}}
        }
        {
        (n_{K_q}-n_{K_{q'}})
        (n_{K_q}-n_{K_{q''}})}
        \ket{K_{q'}}.
    \end{align}
    which are the states we use when comparing the numerically computed orbitals $\ket{\psi_{r}}$ in Fig.~\ref{fig:Overlap__Squred_Perturbation}.
    
    We also note that, with these corrected orbitals, one can also compute the correction to perturbative corrections to the occupation $n_{r}$.Suppose we were able to find the exact orbital states $\ket{\psi_{r}}$, which are by definition eigevectors of $\boldsymbol{F}$. Then taking the expectation value of both sides of Eq.~(\ref{eq:Lyaponov}), we obtain
    \begin{align}
        n_{r}
        =
        \frac{
        \langle \psi_r| \boldsymbol{G} |\psi_{r}\rangle
        }
        {
        \langle \psi_r| \boldsymbol{G} |\psi_{r}\rangle
        +
        \langle \psi_r| \boldsymbol{L} |\psi_{r}\rangle
        }
    \end{align}
    Replacing $\ket{\psi_{r}}$ with its perturbative correction gives the occupation to the desired order. 
    
    \section{Two additional non-Hermitian tight-binding models} \label{app:Two_More_Models}
    \subsection{Non-Hermitian SSH model}
    In this appendix, we will briefly study one version of the non-Hermitian SSH model initially considered in Ref.~\cite{Lieu_PRB_2018}. We will restrict ourselves to the fermionic case, since the analysis for the bosonic case essentially follows the same logic. To obtain a quantum-mechanically consistent effective Hamiltonian, we will follow Method 2 and add a minimal amount of loss. The effective non-Hermitian Hamiltonian reads
    \begin{align}\nonumber
        \hat{H}^{\rm SSH}_{\rm eff}
        =
        &\sum_{j}
        (\frac{w-\kappa}{2} \hc_{A,j}^\dagger \hc_{B,j}
        +
        \frac{w+\kappa}{2} \hc_{B,j}^\dagger \hc_{A,j})
        \\ \nonumber
        +&
        \sum_{j}
        (\frac{u+\gamma}{2} \hc_{A,j+1}^\dagger \hc_{B,j}
        +
        \frac{u-\gamma}{2} \hc_{B,j}^\dagger \hc_{A,j+1})
        \\ \label{eq:SSH_H_eff}
        -i&
        \sum_{j}
        (\Gamma+\frac{\kappa+\gamma}{2})
        (
        \hc_{A,j}^\dagger \hc_{A,j}
        +
        \hc_{B,j}^\dagger \hc_{B,j}
        )
    \end{align}
    where $A$ and $B$ label the two sites in a unit cell, $w\pm\kappa$ are the intra cell hopping and $u\pm \gamma$ the inter cell hopping. This model is known to exhibit the NHSE for any value of non-reciprocal hopping (in the sense that the spectrum collapses to a line in the complex plane). Note that if $u= w$ and $\gamma = \kappa$, we recover the Hatano-Nelson model.
    
    We first wish to describe this model for an infinitely large lattice. Fourier transforming to momentum space we obtain
    \begin{align}
        \hat{H}^{\rm SSH}_{\rm eff}
        =
        \sum_{k}
        \boldsymbol{\hat{C}}_k^\dagger
        \boldsymbol{H}_{\rm eff}(k)
        \boldsymbol{\hat{C}}_k
    \end{align}
    where $\boldsymbol{\hat{C}}_k^\dagger = (\hc_{A,k}^\dagger, \hc_{B,k}^\dagger)$ and
    \begin{align}
    \boldsymbol{H}_{\rm eff}(k)
    =
    \begin{pmatrix}
    -i(\Gamma+\frac{\kappa+\gamma}{2}) & \frac{w-\kappa}{2}+\frac{u+\gamma}{2}e^{-i k}
    \\
    \frac{w+\kappa}{2}+\frac{u-\gamma}{2}e^{i k} & -i(\Gamma+\frac{\kappa+\gamma}{2})
    \end{pmatrix}
    \end{align}
    is the effective Hamiltonian. There are two bands, and the complex energies are
    \begin{align} \nonumber
    E_{\pm}(k)
    &=
    \pm\frac{1}{2}
    \sqrt{
    J^2+\tilde{J}^2
    +2 J \tilde{J}\cos(k+i(A+\tilde{A}))
    }
    \\
    &
    -
    i(\Gamma+\frac{\kappa+\gamma}{2})
    \end{align}
    where $J = \sqrt{w^2-\kappa^2}, 2A = \ln(w+\kappa)/(w-\kappa)$ as before and we have introduced $\tilde{J} = \sqrt{u^2-\gamma^2}, 2\tilde{A} = \ln(u+\gamma)/(u-\gamma)$. Despite the potentially interesting band structure, the steady state is completely determined by the anti-Hermitian part of the effective Hamiltonian. As discussed in the main text, the baths can not cause transitions between momentum eigenstates. In the infinite time limit the only structure that remains in the incoherent pumping and decay. As before, the most interesting feature of this model is the open-boundary steady-state occupation.
    
    We would like to argue that the most interesting feature we see in the Hatano-Nelson model, namely the existence of a large length scale that is directly tied to real-space propagation dynamics, is still present in this generalized model. To this end, we will compare two different methods for computing $\langle \hc_{A/B,j}^\dagger \hc_{A/B,j} \rangle_{\rm ss}$ under open boundary conditions. We first numerically solve the Lyaponov equation Eq.~(\ref{eq:Lyaponov}), which is straightforward. For the second method, we will use the formal solution Eq.~(\ref{eq:Fnm_Formal_Solution}) except we will approximate the open-boundary Green's function by that of an infinite chain (just as we did in App.~\ref{app:Steady_State_Occupation}).
    
    We therefore first need to find the real-space Green's funciton for an infinite-sized lattice. The momentum- space retarded Green's function is simply
    \begin{align}\nonumber
        &\begin{pmatrix}
            G^R_{\infty,AA}[k;\omega] & G^R_{\infty, AB}[k;\omega] \\
            G^R_{\infty, BA}[k; \omega] & G^R_{\infty, BB}[k; \omega]
        \end{pmatrix}
        =
        \frac{1}{\omega \boldsymbol{1}- \boldsymbol{H}_{\rm eff}(k)}
        \\ \nonumber
        &
        =
        \frac{1}{
        [
        (\omega+ i s)^2-\frac{1}{4}(J^2+\tilde{J}^2)-\frac{1}{2}J \tilde{J} \cos(k+i(A+\tilde{A}))
        ]}
        \\
        &
        \times
        \begin{pmatrix}
            \omega + i s & \frac{1}{2}(J e^{-A}+e^{-i k}\tilde{J} e^{\tilde{A}})  
            \\
            \frac{1}{2}(J e^{A}+e^{i k}\tilde{J} e^{-\tilde{A}})   & \omega + i s
        \end{pmatrix}
    \end{align}
    where $s = \Gamma + (\kappa+\gamma)/2$. Taking the Fourier transform to real space, we obtain
    \begin{align}\nonumber
        \begin{pmatrix}
            G^R_{\infty,AA}[j,p;\omega] & G^R_{\infty, AB}[j,p;\omega] \\
            G^R_{\infty, BA}[j,p; \omega] & G^R_{\infty, BB}[j,p; \omega]
        \end{pmatrix}
           =
        \int_{-\pi}^{\pi}
        \frac{d k}{2\pi}
        \frac{e^{i k(j-p)}}{\omega \boldsymbol{1}- \boldsymbol{H}_{\rm eff}(k)}
    \end{align}
    The integral can be computed analytically using the residue theorem. We have
\begin{align}
        &G^R_{\infty, AA}[j,p;\omega]
        =
        \frac{-2 i e^{(j-p)(A+\tilde{A})}}{J \tilde{J} \sin \tilde{Q}[\omega]}
        (\omega+ i s)
        e^{-i \tilde{Q}[\omega]|j-p|}
        \\
        &
        G^R_{\infty, BB}[j,p;\omega]
         =
        \frac{-2 i e^{(j-p)(A+\tilde{A})}}{J \tilde{J} \sin \tilde{Q}[\omega]}
        (\omega+ i s)
        e^{-i \tilde{Q}[\omega]|j-p|}
\end{align}

\begin{align} \nonumber
        G^R_{\infty, AB}[j,p;\omega]
        =
        \frac{- i e^{(j-p)(A+\tilde{A})-A}}{J \tilde{J} \sin \tilde{Q}[\omega]}
        (
         & J e^{-i \tilde{Q}[\omega]|j-p|}
        \\
        +
        &\tilde{J} e^{-i \tilde{Q}[\omega]|j-p-1|}
        )
\end{align}

\begin{align} \nonumber
        G^R_{\infty, BA}[j,p;\omega]
        =
        \frac{- i e^{(j-p)(A+\tilde{A})+A}}{J \tilde{J} \sin \tilde{Q}[\omega]}
        (
         & J e^{-i \tilde{Q}[\omega]|j-p|}
        \\
        +
        &\tilde{J} e^{-i \tilde{Q}[\omega]|j-p+1|}
        )
\end{align}
Using these expressions, we can then approximate the real-space steady-state occupation of this non-Hermitian SSH model as
\begin{align} \nonumber
    \langle \hc_{A,j}^\dagger \hc_{A,j} \rangle_{\rm ss}
    \approx
    2\Gamma
    \sum_{p}
    \int_{-\infty}^{\infty}
    \frac{d \omega}{2\pi}
    (&|G^{R}_{\infty, AA}[j,p;\omega]|^2
    \\ \label{eq:A_Approx_Occupation}
    +
    &
    |G^{R}_{\infty, AB}[j,p;\omega]|^2
    ).
\end{align}
\begin{align} \nonumber
    \langle \hc_{B,j}^\dagger \hc_{B,j} \rangle_{\rm ss}
    \approx
    2\Gamma
    \sum_{p}
    \int_{-\infty}^{\infty}
    \frac{d \omega}{2\pi}
    (&|G^{R}_{\infty, BA}[j,p;\omega]|^2
    \\ \label{eq:B_Approx_Occupation}
    +
    &
    |G^{R}_{\infty, BB}[j,p;\omega]|^2
    ).
\end{align}
by numerically computing the integrals over frequency.

\begin{figure}[t]
	\centering
	\includegraphics[width=0.475\textwidth]{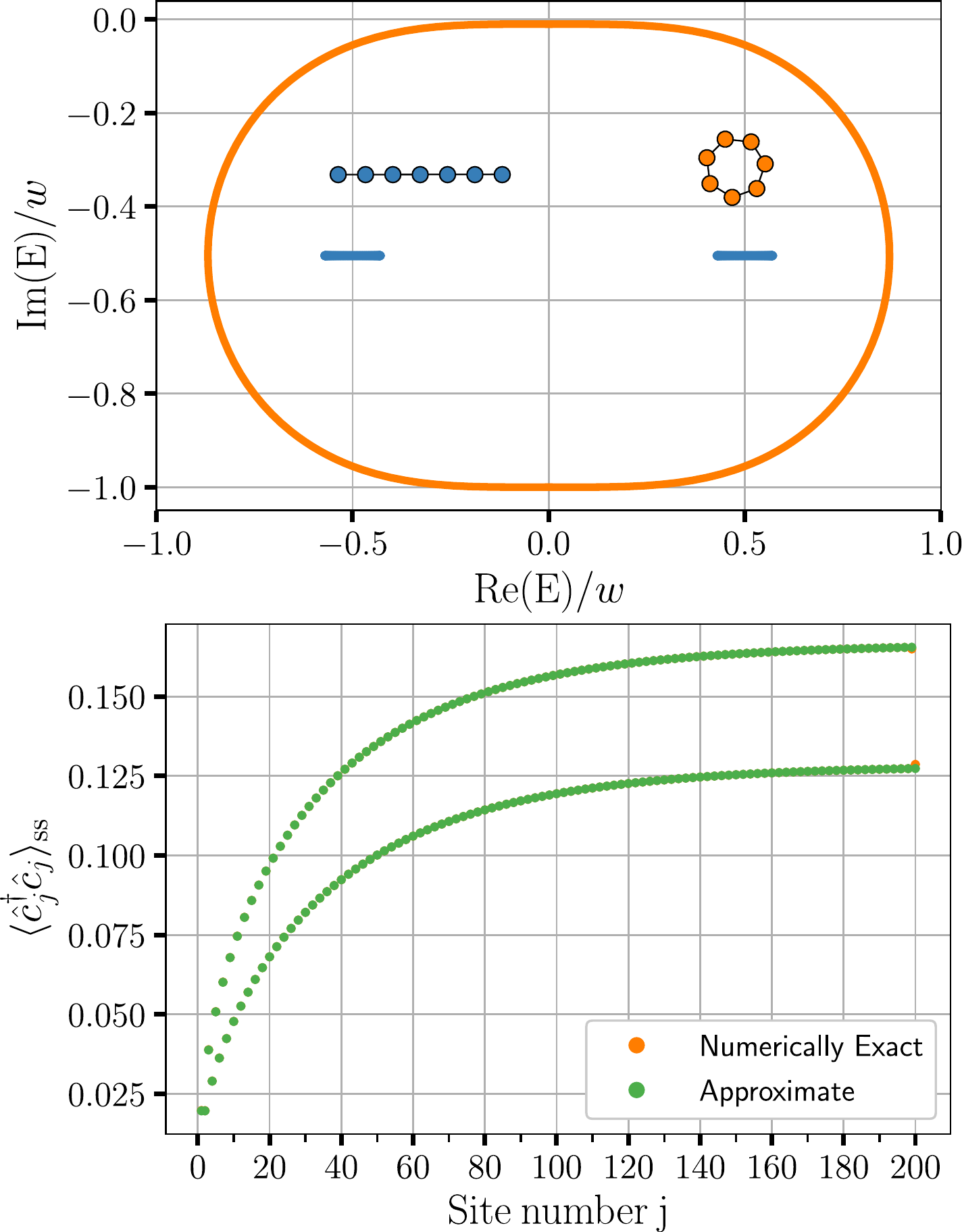}
	\caption{Top: Periodic (orange) and open (blue) boundary spectrum of the SSH model described in Eq.~(\ref{eq:SSH_H_eff}) for $w = u = 1, \kappa = 0, \gamma = 0.99$ and $\Gamma = 0.01$ (i.e. only non-reciprocal hopping on every second bond). The model is known to exhibit the NHSE: all the energies to collapse to the real line and all right eigenvectors are localized to one side of the chain. The periodic boundary system has a gap of order $\Gamma$, whereas the open chain has a gap of order $(\kappa+\gamma)/2$. Bottom: Plot of the real-space steady-state occupation $\langle \hc_j^\dagger \hc_j \rangle_{\rm ss}$ for the parameters above with $N = 200$ total sites (100 unit cells). Even and odd sites correspond to the $A$ and $B$ sublattice degrees of freedom respectively. Despite the large damping gap, there is still a large length scale which characterizes the steady-state occupation.  The numerically exact method, which we obtain by directly solving the matrix equation Eq.~(\ref{eq:Lyaponov}) is essentially identical to the solution obtained by approximating the open-boundary Green's function by its infinite-system counterpart (see Eqs.~(\ref{eq:A_Approx_Occupation}))-(\ref{eq:B_Approx_Occupation}).   }
	\label{fig:SSH_Model}
	\end{figure}

In Fig.~\ref{fig:SSH_Model} we plot the numerically exact result (which comes from directly solving Eq.~(\ref{eq:Lyaponov})), the approximate solution Eqs.~(\ref{eq:A_Approx_Occupation})-(\ref{eq:B_Approx_Occupation}) in addition to the open and periodic spectrum for a given choice of parameters. There are two salient features. First note that there is still a large length scale which describes the steady-state occupation, despite the existence of a dissipative gap of order $(\kappa+\gamma)/2$. The length scale associated with this gap would only be on the order of a few lattice sites, which is evidently not how $\langle \hc^\dagger_j \hc_j \rangle_{\rm ss}$ behaves. Further, seeing as the approximate and exact solutions are nearly identical, it follows that this length scale is encoded in the the Green's function for an infinite system, and has nothing to do with the open-boundary geometry. This once again justifies the approximation that, in the bulk, the retarded Green's function is largely unaffected by the NHSE, even though the eigenvalues and eigenvectors change dramatically depending on boundary conditions. \\

\subsection{Hatano-Nelson with next-nearest neighbour hopping}
\label{app:HN_NNN}

We now analyze another non-Hermitian tight-binding model: the Hatano-Nelson model with an additional next- nearest-neighbour (NNN) Hermitian hopping term. Since the anti-Hermitian part of the effective Hamiltonian is the same as the Hatano-Nelson model we can use the same set of dissipators. The effective Hamiltonian reads
\begin{align} \nonumber
    \hat{H}_{\rm eff}^{NNN}
    &=
    \sum_{j}
    \left(
    \frac{w+\kappa}{2} \hc_{j+1}^\dagger \hc_{j}
    +
    \frac{w-\kappa}{2} \hc_{j}^\dagger \hc_{j+1}
    \right)
    \\ \nonumber
    &
    +
    \frac{T}{2}
    \sum_{j}
    \left(
    e^{i \phi }
    \hc_{j+2} \hc_j
    +
    e^{-i \phi}
    \hc_{j}\hc_{j+2}
    \right)
    \\ \label{eq:H_eff_NNN}
    &
    -i(\Gamma+\kappa)
    \sum_{j}
    \hc_j^\dagger \hc_j
\end{align}
where $T$ is the real hopping amplitude and $\phi$ an arbitrary real phase. Note that if $\phi \neq 0, \pi$ we have broken time-reversal symmetry, as the phase $\phi$ cannot be gauged away. We once again stress that for periodic boundary conditions, only the dissipation determines the steady state which is characterized by the momentum-space occupation, see Eq.~(\ref{eq:Momentum_Occupation_SS}). We therefore only consider the open boundary case.

\begin{figure}[t]
	\centering
	\includegraphics[width=0.475\textwidth]{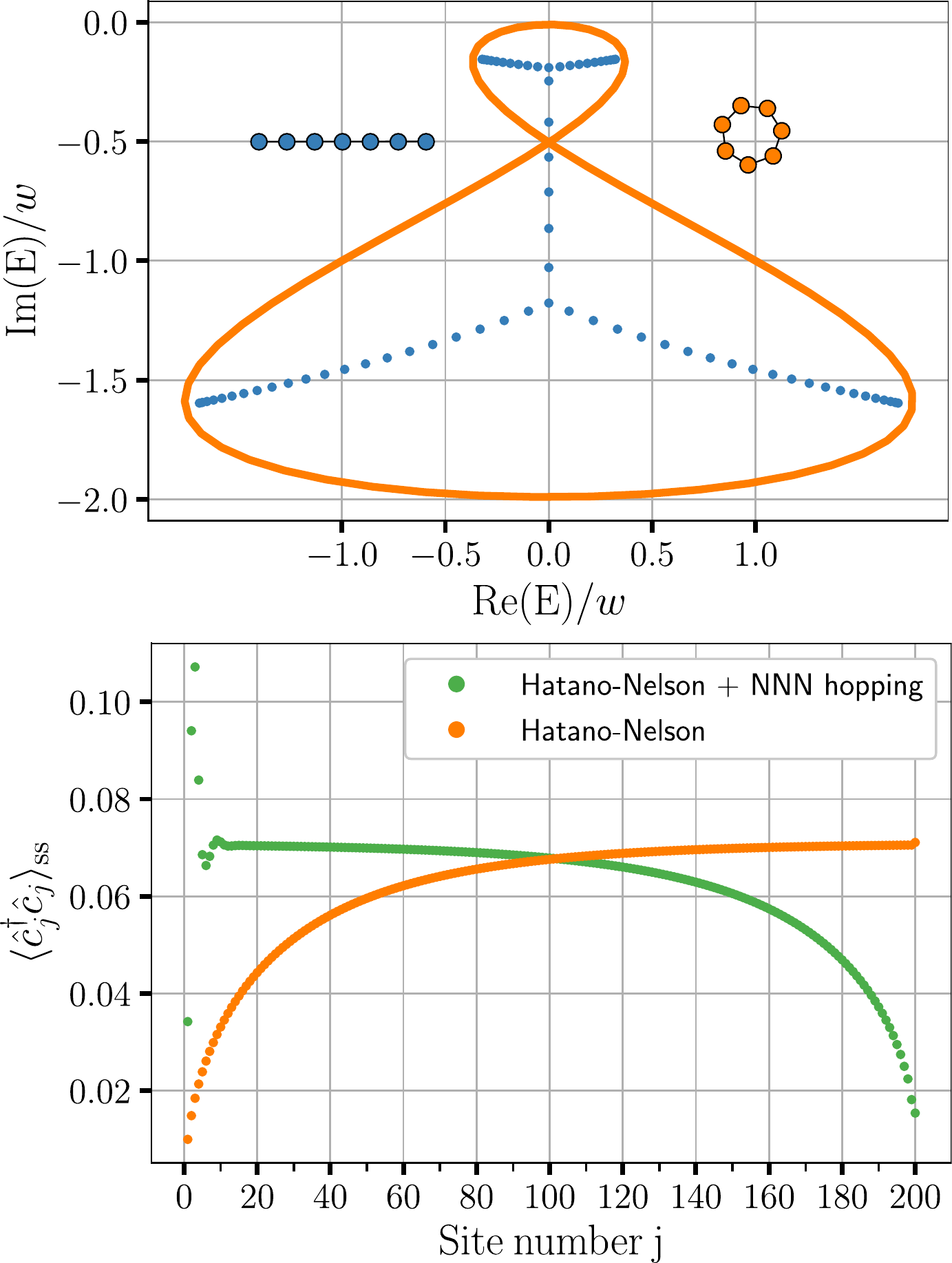}
	\caption{
	Top:  Periodic (orange) and open (blue) boundary spectrum of the Hatano-Nelson model with NNN hopping described in Eq.~(\ref{eq:H_eff_NNN}) for $w = T = 1$, $\kappa = 0.99$, $\phi = \pi/2$ and $\Gamma = 0.01$. Note that the open chain spectrum is only using $N = 70$ sites, due to numerical stability issues in computing the eigenvalues. The periodic boundary system has a gap of order $\Gamma$, whereas the open chain gap is of order of magnitude smaller than $\kappa$. The system exhibits the NHSE, with nearly 60\% of all right eigenvectors localized to the right, and 40\% localized to the left of the chain under open boundary conditions. Bottom: Plot of the real-space steady-state occupation $\langle \hc_j^\dagger \hc_j \rangle_{\rm ss}$ for the Hatano-Nelson model and the Hatano-Nelson model with NNN hopping with the parameters above for a chain with $N = 200$ sites. The two models share the same large length scale $\xi_{\rm obc} \approx w/(2\Gamma)$ despite having very different eigenvalues and eigenvectors.  
	}
	\label{fig:NNN_Hopping}
	\end{figure}

We know that if the next-nearest-neighbour hopping is zero, the propagation dynamics favors rightward propagation. The naive assumption is that this still holds for arbitrary $T$ and $\phi$, since 
the added next-nearest-neighbour hopping
does not \textit{a priori} favor left or right propagation
as it is Hermitian. Using this line of thinking, we would expect particles to pile up on the right side, just like the original Hatano-Nelson model. In Fig.~\ref{fig:NNN_Hopping} we show that there exists a set of parameters for which the opposite is true: in the steady state particles pile up on the left side, but with the same long length scale $\xi_{\rm obc} \approx w/(2\Gamma)$ as in the Hatano-Nelson model. There is still a dissipative gap, and thus the naive expectation is that the associated length scale is very small. Further, the periodic and open boudnary spectrum looks nothing like the original Hatnao-Nelson model. How are we to understand this behavior? 

The answer is to consider the dispersion of an infinite-sized chain, which can be readily found from Eq.~(\ref{eq:H_eff_NNN}) and gives $E(k) = w \cos (k) + T \cos(2k- \phi)-i \kappa(1+ \sin k)-i \Gamma$. The least-damped mode is at $k = -\pi/2$, and the corresponding group velocity for the chosen parameters $T = w$ and $\phi = \pi/2$ is $w \partial_k (\cos k)|_{k = -\pi/2} + w \partial_k \cos(2 k - \pi/2) |_{k = -\pi/2}  = - w < 0$. Thus, with the inclusion of the next-nearest neighbour hopping, the least-damped mode now propagates to the left instead of the right. Further, we can once again estimate the relevant decay length as the group velocity divided by the residual decay rate $\Gamma$ and recover $\xi_{\rm obc} \approx w/(2\Gamma)$.  Note that the right and left eigenvectors of the open chain also resemble nothing like the steady-state occupation $\langle \hc_j^\dagger \hc_j \rangle_{\rm ss}$. With the chosen parameters, about 60\% are localized to the right of the chain, and nearly 40\% are localized to the left.  \\

    \section{Steady state for 
    noise with real-space correlations
    }\label{app:G_Not_Identity}
    Here we briefly study a model whose noise matrix $\boldsymbol{G}$ has a non-trivial spatial correlations. As we have shown in the main text, non-reciprocal dynamics can lead to an interesting steady state even when the noise is uniform. To disentangle the effects of non-reciprocity and the possibly interesting consequences of real-space fluctuation correlations, we seek a set of dissipators and coherent Hamiltonian which give rise to an effective reciprocal Hamiltonian. Working with fermionic particles, this can be achieved by choosing      
    \begin{align} \label{eq:H_coh_recip}
			\hH 
			&= 
			\frac{w}{2}
			\sum_{j}
			\left(
			\hc_{j+1}^\dagger\hc_j+h.c.  
			\right)
			\\ \label{eq:L_n_recip}
			\hat{L}_j &= 
			\sqrt{\kappa} \left(\hc_j-i \hc_{j+1}\right)
			\\ \label{eq:G_n_recip}
			\hat{G}_j &= 
			\sqrt{\Gamma} \left(\hc_j^\dagger-i \hc_{j+1}^\dagger\right)
	\end{align}
	and setting the decay and pumping rate to be the same $\kappa = \Gamma$. The effective Hamiltonian and noise matrix are
	\begin{align}
	    \boldsymbol{H}_{\rm eff}
	    &=
	    \boldsymbol{H}-i 2 \Gamma \boldsymbol{1}
	    \\
	    \boldsymbol{G}
	    &=
	    -i\Gamma
	    \sum_{j}
	    \left(
	    \ket{j+1}\bra{j}
	    -
	    \ket{j}\bra{j+1}
	    \right)
	    +
	    2\Gamma \boldsymbol{1}
	\end{align}
	The effective Hamiltonian corresponds to a reciprocal tight-binding model with a uniform decay rate $\Gamma$. The noise matrix also has a similar structure, except the hopping matrix element is purely imaginary. This leads to an incompatibility between the dynamics and the noise, which can formally be written as $[\boldsymbol{H}, \boldsymbol{G}] \neq 0$: $\boldsymbol{H}$ and $\boldsymbol{G}$ can not be diagonalized by a common set of eigenvectors. Consequently, the dynamics can cause transitions between particles added by the gain baths to any eigenstate of $\boldsymbol{G}$ as is made clear by the formal solution to the steady-state correlation matrix $\boldsymbol{F}$ in Eq.~(\ref{eq:F_Right_Left}).
	
	\begin{figure}[t]
	\centering
	\includegraphics[width=0.475\textwidth]{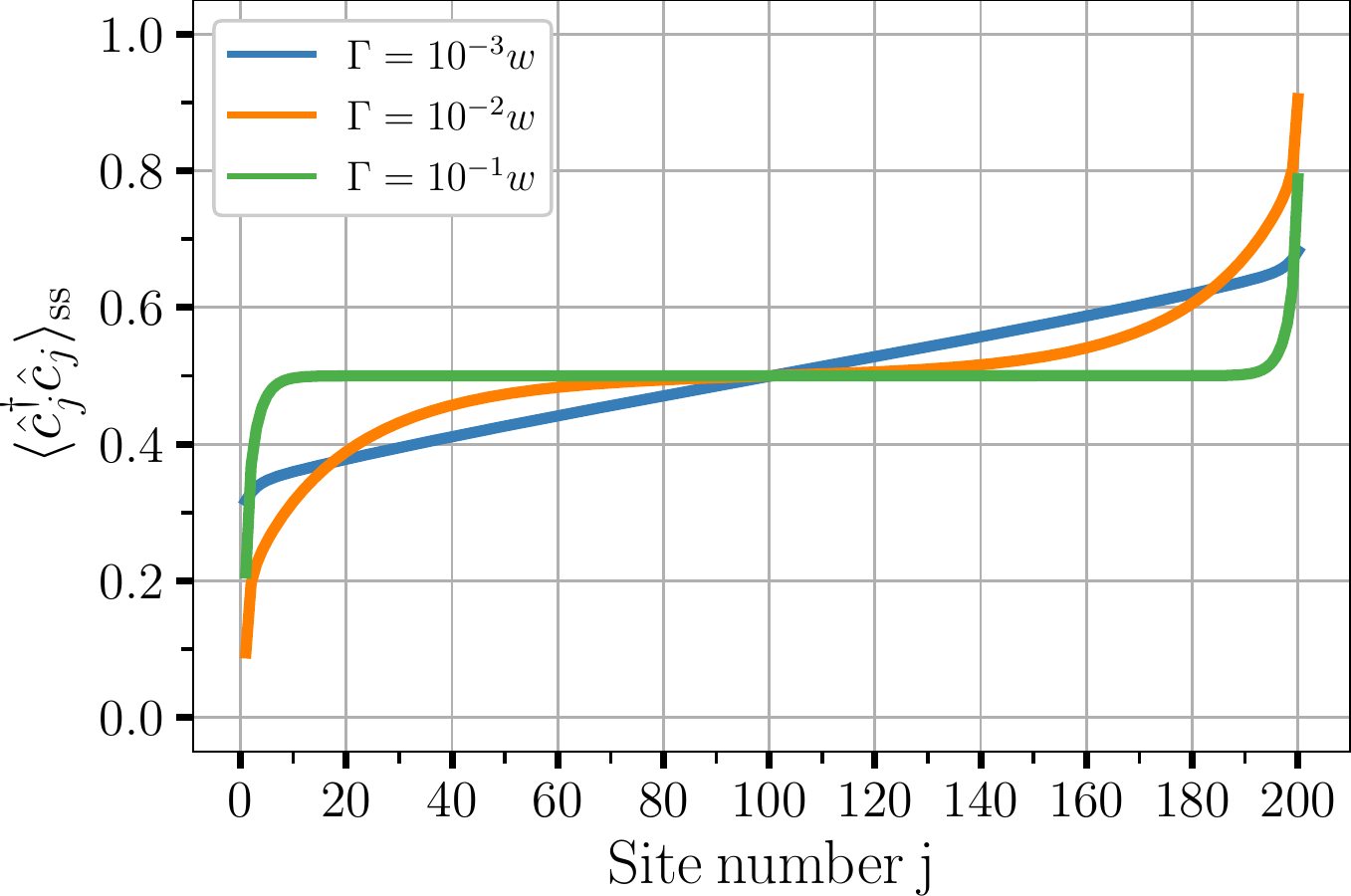}
	\caption{Steady-state occupation $\langle \hc_j^\dagger \hc_j \rangle_{\rm ss}$ of a reciprocal tight-binding model with a non-trivial noise correlation matrix. The effective Hamiltonian and noise are realized with the coherent Hamiltonian and dissipators Eqs.~(\ref{eq:H_coh_recip}-\ref{eq:G_n_recip}). Despite the lack of non-reciprocity, there is an accumulation and depletion of particles on opposite ends of the chain.  This can be attributed to non-uniform pumping, which adds right-movers to the chain at a higher rate than left-movers. }
	\label{fig:NonTrivial_G}
	\end{figure}
    
	The discord between the noise and the dynamics leads to a non-trivial steady-state occupation $\langle \hc_j^\dagger \hc_j \rangle_{\rm ss}$ as we show in Fig.~\ref{fig:NonTrivial_G}. In the bulk, we recover the expected occupation of half-filling, seeing as we have set $\kappa = \Gamma$. The occupation at the boundaries however is non-trivial, with an excess of particles at one edge and an equivalently depleted number at the opposite edge. We stress that this effect can not be attributed to non-reciprocity, as there is none. In a similar vain, it can not be explained by any novel non-Hermitian phenomena such as the NHSE: the dynamical matrix $\boldsymbol{H}$ is only trivially non-Hermitian in that it has a uniform decay.
	
	The goal of this appendix is not to fully characterize the steady state of this model, but rather to point out how structured fluctuations can lead to interesting behavior even when the dynamics are reciprocal.  That being said, we can provide a simple intuitive reason as to why there is an accumulation of particles on one edge and a lack of them on the other. We first note that the eigenvectors of $\boldsymbol{G}$ are standing-wave states with a center of mass momentum $\pi/2$
	\begin{align}\label{eq:Eigen}
	    \braket{j}{K_{q}}
	    =
	    \sqrt{\frac{2}{N+1}}e^{i \frac{\pi}{2}j}\sin K_q j
	\end{align}
	with corresponding eigenvalues
	\begin{align}\label{eq:G_Eigenvalues}
	    G_{K_q}
	    =
	    2\Gamma
	    (
	    1-\cos K_q
	    ).
	\end{align}
    Using the dispersion our system, Eq.~(\ref{eq:Eigen}) tells us that the group velocity of a standing wave with momentum $K_q$ is $\partial_k (w \cos k) |_{k= \frac{\pi}{2} \pm K_q } = -w \cos K_q$. Further, recall that the eigenvalues of $\boldsymbol{G}$ correspond to the rate at which the baths add particles to the corresponding eigenstate. Together with Eq.~(\ref{eq:G_Eigenvalues}), we thus see that the baths adds right-moving particles $-w\cos K_q >0$ at a higher than than left moving particles $-w \cos K_q < 0$. This succinctly explains why, at least qualitatively, there is an population imbalance in the steady state. The only dimensionless parameter in the problem, $w/\Gamma$, also controls a length scale which determines how the boundary occupation deviates from half-filling. As we increase $\Gamma$ particles decay more quickly out of the system, and the number disparity between left-movers and right-movers because immaterial. 
    
    
    \bibliography{Dissipative_HN_Bib}
	
 \end{document}